\documentclass[11pt]{article}
\usepackage{amsfonts}
\usepackage{amssymb}
\usepackage{amsmath}
\usepackage{soul}
\usepackage{amsthm}
\usepackage{epsfig}
\usepackage{graphicx}
\usepackage{bm}
\usepackage{color}
\usepackage{hyperref}
\usepackage{hyperref,breakurl,amsmath,amssymb,pst-node,eepic}
\usepackage{verbatim}
\usepackage{todonotes}

\newlength{\bredde}
\def\slash#1{\settowidth{\bredde}{$#1$}\ifmmode\,\raisebox{.15ex}{/}
\hspace*{-\bredde} #1\else$\,\raisebox{.15ex}{/}\hspace*{-\bredde} #1$\fi}
\newcommand{\be}{\begin{equation}}
\newcommand{\ee}{\end{equation}}
\newcommand{\bea}{\begin{eqnarray}}
\newcommand{\eea}{\end{eqnarray}}

\newcommand{\eins}{\leavevmode\hbox{\small1\kern-3.8pt\normalsize1}}
\newcommand{\e}{\mbox{e}}

\newcommand{\Tr}{\mbox{Tr}}
\newcommand{\lla}{\left\langle}
\newcommand{\rra}{\right\rangle}
\newcommand{\sect}[1]{\setcounter{equation}{0}\section{#1}}
\renewcommand{\theequation}{\thesection.\arabic{equation}}

\date{\today}

\newcommand{\conj}[1]{\overline{#1}}

\newcommand{\diff}{\textup{d}}

\renewcommand{\imath}{\textup{i}}
\renewcommand{\Re}{\mathfrak{R}}
\renewcommand{\Im}{\mathfrak{I}}

\begin{document}
\topmargin -1.4cm
\oddsidemargin -0.8cm
\evensidemargin -0.8cm
\title{\Large{{\bf
Universal eigenvector correlations in quaternionic Ginibre ensembles 
}}}

\vspace{1.5cm}

\author{~\\{\sc Gernot Akemann}$^{1,2}$, {\sc Yanik-Pascal F\"orster}$^{3}$, and
{\sc Mario Kieburg}$^{4}$
\\~\\
$^1$Faculty of Physics,
Bielefeld University,\\
P.O. Box 100131,
33501 Bielefeld, Germany\\~\\
$^2$Department of Mathematics, Royal Institute of Technology (KTH),\\ 
Brinellv\"agen 8, 114 28 Stockholm, Sweden
\\~\\
$^3$
Department of Mathematics, King’s College London, \\ 
Strand, London WC2R 2LS, United Kingdom\\~\\
$^4$School of Mathematics and Statistics,
The University of Melbourne,\\
Parkville, VIC 3010, Australia
}

\maketitle
\vfill
\begin{abstract}

Non-Hermitian random matrices enjoy non-trivial correlations in the statistics of their eigenvectors. We study the overlap among left and right eigenvectors in Ginibre ensembles with quaternion valued Gaussian matrix elements. This concept was introduced by Chalker and Mehlig in the complex Ginibre ensemble.  Using a Schur decomposition, for harmonic potentials we can express the overlap in terms of complex eigenvalues only, coming in conjugate pairs in this symmetry class. Its expectation value leads to a Pfaffian determinant, for which we explicitly compute the matrix elements for the induced Ginibre ensemble with $2\alpha$ zero eigenvalues, for finite matrix size $N$. In the macroscopic large-$N$ limit in the bulk of the spectrum we recover the limiting expressions of the complex Ginibre ensemble for the diagonal and off-diagonal overlap, which are thus universal.

\end{abstract}
\vfill

\thispagestyle{empty}
\newpage

\renewcommand{\thefootnote}{\arabic{footnote}}
\setcounter{footnote}{0}

\sect{Introduction}

The analysis of spectral statistics in random matrix theory (RMT) and the comparison of its universal predictions to spectral data from physics (or other sciences) has a long tradition, see \cite{Mehta,Guhr1998,Peter,ABF} for references including applications. Compared to that the statistics of eigenvectors has received less attention. This is probably because the eigenvectors of the standard 
complex 
Hermitian random matrices are distributed according to the Haar measure on the unitary group which is known to be universal, cf. \cite{Knowles} and references therein. This is in contrast to the eigenvalues in RMT, being strongly correlated random variables. The situation changes dramatically when considering non-normal Hamilton operators with complex spectra. Motivated by the strong sensitivity of eigenvectors of such operators to perturbations, Chalker and Mehlig \cite{CM,CM:2000} introduced the overlap between left and right eigenvectors in non-Hermitian RMT, as the left (or right) eigenvectors no longer form an orthonormal system by themselves.

There are many physics applications of complex non-Hermitian RMT in general, including open quantum systems, diffusion or dynamics in random media, see e.g. \cite{FS} for a review. Specifically the symmetry class of quaternionic Ginibre matrices   \cite{Ginibre1965} considered here
represents a static two-dimensional Coulomb gas with constant background charge at fixed inverse temperature $\beta=2$ \cite{Peter}, which is of interest as a statistical mechanics system in its own right.
Furthermore, it has applications to Hamiltonians with random potential and imaginary magnetic field \cite{Efetov}, as studied in the context of vortices in superconductors. A map of this ensemble to a fermionic field theory has been suggested in \cite{Hastings2000}, and a chiral version of this symmetry class has been successfully compared to complex spectra from the QCD Dirac operator with two colours and chemical potential \cite{AB}. We expect that our results for the overlap in quaternionic Ginibre ensembles apply in all these settings
in the bulk of the spectrum.

Recently, there has been much progress in determining the correlations of overlaps of eigenvectors in RMT, see \cite{WS,Fyodorov2018,FGS,Bourgade2018,CR,BGZ,ATTZ,FT} for a non-exhaustive list. Many of these focus on the complex and real Ginibre ensemble 
introduced in  \cite{Ginibre1965}. 
Further symmetry classes such as products of Ginibre ensembles  \cite{PZ} or spherical and truncated ensembles \cite{GD2} have been studied as well.
The question of universality of eigenvector statistics is so far not fully understood, cf. \cite{MANT} summarising nicely the above results and formulating their conjectured universality. 
We will thus investigate the third, quaternionic Ginibre ensemble \cite{Ginibre1965,Mehta} in order to compare to the above classes, see also \cite{BenaychGeorges2011,GD} for earlier studies. It has become 
apparent only quite recently 
that all three Ginibre ensembles share the same local eigenvalue  statistics  in the bulk  \cite{BS,AKMP} and at the edge of the spectrum \cite{Rider,BS}. One may therefore expect that the same holds true for the local eigenvector statistics in the bulk and at the edge, see \cite{ATTZ} for the corresponding results in the complex Ginibre ensemble. The edge statistics is particularly relevant for scattering in chaotic cavities, cf. 
\cite{FSPB,SFPB}.
 
Let us give a very brief overview over the most recent developments in the computation of eigenvector statistics. Initiated by Chalker and Mehlig \cite{CM,CM:2000}, many authors have studied the average of the overlap matrix, consisting of the scalar product between two left times that of two right eigenvectors. In case these belong to the same complex eigenvalue this matrix is called diagonal overlap, else it is called off-diagonal overlap.
An important ingredient in applications of RMT is the introduction of a time dependence or dynamics on the set of eigenvalues, leading to Dyson's Brownian motion for the classical Hermitian ensembles. In a series of papers, using Green's functions and stochastic evolution equations, it was emphasised that for non-Hermitian random matrices the eigenvector and eigenvalue dynamics no longer decouples \cite{Burda1,Burda2,MANT}, with references to further applications e.g. to neural networks.
This aspect of coupling was raised much earlier in \cite{SavinSokolov}, in terms of the so-called Bell-Steinberger nonorthogonality matrix.

Rigorous results have been obtained for the distribution of diagonal and off-diagonal overlaps and their correlations for the complex Ginibre ensemble (GinUE) in \cite{Bourgade2018} using probabilistic,  and  in \cite{Fyodorov2018,FGS} using supersymmetric and orthogonal polynomial techniques,  respectively. The latter also included partial results on the real eigenvalues of the real Ginibre ensemble (GinOE).  
The authors of \cite{Bourgade2018} numerically assessed the universality of their results in some examples.
The correlation of angles between eigenvectors in the GinUE was analysed in \cite{BGZ}. Starting from moments of the overlap matrix \cite{WS,CR}, a determinantal structure in terms of a kernel was derived in \cite{ATTZ} for the conditional overlaps for finite matrix size $N$, that 
allowed to take various large-$N$ limits. The question of localisation of eigenvectors in non-Hermitian RMT was answered in  \cite{Rudelson,O'Rourke}, going beyond Gaussian ensembles, but we will not pursue this direction in our quaternionic Ginibre ensemble (GinSE).

In \cite{YanDima} a parametric width
velocity was derived as an indicator of
nonorthogonality and was verified later in
microwave cavity and reverberation chamber experiments \cite{Gros}.
Another way to asses the stability of eigenvectors under perturbation is through the condition number, that was studied from free probability theory in \cite{Speicher} and very recently for the GinOE in \cite{FT}.

The results presented here are based on \cite{YPF} developed in parallel to the publication \cite{GD}. Whereas \cite{GD} uses powerful techniques from probability theory to derive rigorous results for the diagonal overlap, we have first focused on results for finite 
matrix dimension, both for the diagonal and off-diagonal overlap. Second, our heuristic global asymptotic results hold for both overlaps in the bulk of the spectrum and thus complement the results of \cite{GD}.

Our findings in this paper are organised as follows. In Section \ref{setup}
we recall the definition of quaternionic random matrices, introduce the ensembles we will study and define overlaps between eigenvectors, distinguishing eigenvalues and their complex conjugate partners. The different overlap matrices are averaged over eigenvalues and an upper triangular matrix $T$, resulting from a Schur decomposition. Section \ref{sec:inductive_calculation} is devoted to the exact determination of the overlaps at finite matrix size $N$, by computing these averages. Here, the expectation with respect to $T$ can be computed for general harmonic potentials, including the elliptic Ginibre ensemble. The remaining average leads to a Pfaffian determinant that is explicitly  computed for the induced quaternionic Ginibre ensemble that has a fixed number of extra zero modes, cf. Appendix \ref{appA} for details. In Section \ref{largeN} we take the large-$N$ limit, first on a macroscopic scale in the bulk of the spectrum. It agrees with the results for the GinUE and is thus universal. We then investigate the local eigenvector statistics in the vicinity of the origin which is specific for this symplectic symmetry class, showing repulsion off the real axis.  
Our conclusions and open questions are presented in Section \ref{conclusion}.

\sect{Eigenvectors in quaternionic random matrix ensembles}\label{setup}

\subsection{Notation and definition of overlap matrix}

In this subsection, we introduce our notation, recalling some facts about quaternions, and define the overlap.
	In the following we will use a standard complex embedding  $ \chi(G) \in M_{2N}(\mathbb{C}) $ of a matrix with quaternion elements $ G\in M_N(\mathbb{H}) $: 
	\begin{align}
\label{eq:embedding_quaternion_to_complex} 
	a+\hat{i}
 a_1 + \hat{j}b_1 +\hat{k}b_2	
	&\mapsto \begin{pmatrix}  a & b \\ -\conj{b} & \conj{a} \end{pmatrix} \quad \text{with}\  a=a_1+\imath a_2,\ b=b_1+\imath b_2 \in \mathbb{C}\ ,	
	\end{align} 
	applied on each matrix entry. 
	Here, $ \hat{i}$, $\hat{j} $ and $ \hat{k} $ are the quaternion units, $\imath$ the imaginary unit, and $a_1,a_2,b_1,b_2\in\mathbb{R}$. {Complex conjugation} for a complex number $ z=x+\imath y $ is given by $\conj{z}= x-\imath y$, and 
	$ \Re(z) = x $ and $ \Im(z) = y $ denote the real and imaginary part of $ z $. For matrices and vectors with components in  $ \mathbb{C} $ or $ \mathbb{H}$, we write $ G^\dagger := \conj{G}^T $ for the {(Hermitian) adjoint} of $ G $, at which $ G^T $ is the {transpose} of $ G $.
	
	Generally, a matrix $ G $ with quaternion elements has a set of infinitely many quaternion (left or right) eigenvalues $ p $, due to a possible similarity transformation $ q^{-1}pq $, where $ q $ is a non-zero quaternion. After introducing  equivalence classes $ [p] $ these contain exactly two complex numbers, that are complex conjugate to each other (if $ p $ is non-real).  In the following we will use these complex, so called {standard} eigenvalues, because their image under $ \chi $ is diagonal. The fact that a complex $ 2N $ dimensional representation $G$ of an 
$ N $ dimensional matrix with quaternion elements has $ 2N $ eigenvalues coming in complex conjugate pairs is often referred to as an aspect of Kramers degeneracy.

In the following we will require two properties to be satisfied: 
Each of the eigenvalues  must belong to one of the eigenvalues of a complex conjugate pair 
and both eigenvalues have to be non-degenerate, that is non-real. For the ensembles of random matrices to be considered below this will only be violated on a set of measure zero.

Under these conditions on $G$ 	it follows for a given right eigenvector $R\in\mathbb{C}^{2N}$ with an eigenvalue $ \lambda_R\in \mathbb{C}$, i.e. 
\begin{equation}
\label{Rev}
GR=\lambda_R R \ ,
\end{equation}
that there is a second right eigenvector $R^\prime= \hat{\tau}_2\conj{R}$ of $G$ 
with eigenvalue $\conj{\lambda}_R$
(see e.g. \cite[Lem.~2.3]{Loring}\footnote{The embedding  \eqref{eq:embedding_quaternion_to_complex} is a modification of the one in \cite{Loring}, which does not alter the stated properties. }). 
We employ the $2N\times 2N$ matrix 
$\hat{\tau}_2 = \tau_2\otimes I_N$ with $\tau_2$ being the second Pauli matrix and $I_N$ the $N$-dimensional identity matrix.
It follows that $ R $ and $ R^\prime $ are orthogonal:
		\begin{align}\label{RRorth}
		R^\dagger \cdot R^\prime = (R^\prime)^\dagger \cdot R = 0\ ,
		\end{align}
where we explicitly denote scalar products by $\cdot$\ .
Since Hermitian adjoints of left eigenvectors of $G$ are right eigenvectors of $G^\dagger $, one can see that 
the spectrum of left eigenvalues $\lambda_L$ 
\begin{equation}
\label{Lev}
L^\dag G=\lambda_L L^\dag\ ,
\end{equation}
agrees with the one of the right eigenvalues $\lambda_R$. Furthermore, we conclude that $ (L^\prime){}^\dagger = \conj{L^\prime}^\dagger\hat{\tau}_2 $ is the left eigenvector with eigenvalue $ \conj{\lambda}_L$, and that it is orthogonal to $L$
\begin{align}\label{LLorth}
L^\dagger \cdot L^\prime = (L^\prime)^\dagger \cdot L = 0\ .
\end{align}
If $ G $ is Hermitian, we recover the well known situation in which left and right eigenvectors coincide and form a complete orthonormal system. This is not the case for non-Hermitian matrices. Provided that the spectrum of $ G $ is non-degenerate, {left} and {right} eigenvectors form a {bi-orthogonal} system. Namely, by enumerating left eigenvectors by $ L_i^\dag $, right eigenvectors by $ R_i $ and their respective eigenvalues $ \lambda_i $, $i=1,\ldots,2N$, by the same index, we obtain \cite{AW}
	\begin{align} \label{eq:bi_orthogonality}
	L_i^\dagger \cdot R_j = \delta_{ij},\quad i,j=1,\ldots,2N\ ,
	\end{align} 
	where $ \delta_{ij} $ is the Kronecker delta.
	
Following Chalker and Mehlig \cite{CM:2000}, we 
define the following matrix of 
{(non-orthogonality) overlaps} also called {Chalker-Mehlig-correlators}
by the components 
	\begin{align}
	\mathcal{O}_{ij} := L_i^\dagger\cdot L_j\ R_j^\dagger\cdot R_i \ ,\quad i,j=1,\ldots,2N\ . \label{eq:overlap_def}
	\end{align} 
For the diagonal overlaps $ \mathcal{O}_{ii} $ the term {self-overlap} is also used~\cite{Fyodorov2018}, for $i\neq j$ the $ \mathcal{O}_{ij} $ are called off-diagonal overlaps. As in \cite{CM:2000}
 the components $\mathcal{O}_{ij}$ are invariant under the transformation 
	\begin{align}\label{eq:trafo}
	R_i &\to c R_i, \\ \nonumber
	L_i &\to \frac{1}{c} L_i,
	\end{align}  
	for any non-zero $ c\in\mathbb{C} $ and any $ 1\leq i \leq 2N $. Furthermore, the overlap matrix is invariant under unitary symplectic transformations of the eigenvectors, as these preserve the quaternionic scalar product $ x^\dagger\cdot y $ \cite[Sec.~1.2.4]{Hall2015}.
Note also that the overlap matrix $ \mathcal{O}_{ij}$ is Hermitian, as mentioned in \cite{GD}.

Because the overlaps $\mathcal{O}_{ij}$ depend on all $N$ complex conjugated pairs we introduce the following notation. 
We assign to each eigenvector of a conjugate eigenvalue of our list, say $ \conj{z_i} $, the index $ \conj{i} $:
	\begin{align}\label{eq:bar-notation}
GR_i&=z_iR_i\ ,\quad GR_{\conj i}=\conj{z_i}R_{\conj i}\ , \\
L^\dag_iG&=z_iL^\dag_i\ ,\quad L^\dag_{\conj i}G=\conj{z_i}L^\dag_{\conj i}\ , 
\nonumber
	\end{align}
for all $i=1,\ldots,N$, 
with left and right eigenvectors $L_{\conj{i}}^\dag$ and $R_{\conj{i}}$, respectively.	The overlap is labelled accordingly, e.g. 
\begin{equation}
\label{Obars}
\mathcal{O}_{i\,\conj j}=L_i^\dagger\cdot L_{\conj j}\ R_{\conj j}^\dagger\cdot R_i\ ,\quad i,j=1,\ldots,N\ ,
\end{equation}
and likewise $\mathcal{O}_{\conj i j}$, $\mathcal{O}_{\conj i \conj j}$ and $\mathcal{O}_{ i j}$. 
From the orthogonality of eigenvectors of complex conjugate pairs \eqref{RRorth} and \eqref{LLorth}, $R_{\conj i}^\dag\cdot R_i=0=L_i^\dag\cdot L_{\conj{i}}$ we immediately have that 
\begin{equation}\label{eq:Oibi}
\mathcal{O}_{i\, \conj i}=0\quad \forall i=1,2,\ldots,N\ ,
\end{equation}
as noted already in \cite{GD}.
As in the complex case \cite{CM:2000}, due to the bi-orthogonality \eqref{eq:bi_orthogonality}  that holds for indices with or without conjugation, 
we have the following representations of the identity matrix $I_{N}$ as a dyadic product:
\begin{eqnarray}\label{eq:closure2}
&&\sum_{j=1}^{N} L_j\ R_j^\dag = I_{N}\ ,\quad \sum_{j=1}^{N} L_{\conj j}\ R_{\conj j}^\dag = I_{N}\ ,
\end{eqnarray}
which is also called closure relation. It leads to the identities
\begin{eqnarray}
\label{eq:Oid2}
&&\sum_{j=1}^{N} \mathcal{O}_{ij}=1\ , 
\quad \sum_{j=1}^{N} \mathcal{O}_{\conj{ij}}=1\ , 
\quad i=1,\ldots,N\ ,
\end{eqnarray}
separately on the set of $N$ complex eigenvalues and of complex conjugated ones.

\subsection{Quaternionic random matrix ensembles and mean overlaps}\label{QRMT}

In this subsection we present the random matrix ensembles to be considered. 
The joint probability density $P(G)$ of matrix elements is defined as 
	\begin{equation}
	\label{eq:GinW}
	P(G)\diff [G]
	:= 
C_N	\exp\left[ -{\Tr\, W(G,G^\dag)}\right] 
	\prod_{i,j=1}^{2N} \diff \Re{G_{ij}}\diff \Im{G_{ij}}, 
	\end{equation}
	where $ G \in M_{2N}(\mathbb{C})$ is a matrix with quaternion elements in standard complex representation, $W$ the potential
	and $C_N$ an appropriate normalisation constant. Expectation values of $Q(G)$ are defined as 
	\begin{equation}
	\left\langle\ Q(G)\ \right\rangle :=\int Q(G)\ 	P(G)\ \diff [G]\ .
	\label{vev}
	\end{equation}
In the case when $W(z,\conj{z})=a|z|^2+ V(z)+\conj{V(z)}$,	with $a>0$ and $V(z)$ an analytic function, the potential is called quasi-harmonic, cf. \cite{Zabrodin2006} for the case of complex matrices, where also issues of convergence are discussed. In this case we will be able to express the expectation values of the conditional overlaps, to be defined in \eqref{eq:onepoint} and \eqref{eq:twopoint} below, in terms of the complex eigenvalues only. A prominent example is given by 
\begin{equation}\label{eq:Ginell}
W(z,\conj{z})=\frac{\sigma^{-2}}{1-\tau^2}\left(|z|^2-\frac{\tau}{2}(z^2+\conj{z}^2)\right)\ ,\quad 
C_N=\left(\frac{\sigma^{-2}}{\pi\sqrt{1-\tau^2}}\right)^{2N^2}\ ,
\end{equation}
the elliptic Ginibre (or Ginibre-Girko) ensemble with $\tau\in[0,1)$ \cite{Girko,SCS}. For $\tau=0$ it reduces to the Ginibre ensemble with variance $\sigma^2/2$.
Below we will mainly consider the induced Ginibre ensemble 
\cite{Fischman}
\begin{equation}
\label{eq:Ginind}
W(z,\conj{z})=\sigma^{-2}|z|^2-2\alpha\ln|z|\ ,\quad 
C_N=(\pi \Gamma(\alpha+1)\sigma^{2\alpha+2})^{-2N^2},
\end{equation}
with $\alpha>-1$. For $\alpha\in\mathbb{N}$ it can be obtained from a Ginibre ensemble of rectangular matrices $G$ of size $2N\times(2N+2\alpha)$, with $\alpha$ counting the number of zero-eigenvalue pairs, cf. \cite{Fischman} for the complex Ginibre case. 
Because the potential of the induced Ginibre ensemble \eqref{eq:Ginind} is rotationally invariant, we will be able to explicitly compute the expected conditional overlaps for finite matrix size.

The steps  to arrive at a joint density of complex eigenvalues (and other degrees of freedom), starting from \eqref{eq:GinW}, are well known \cite{Ginibre1965}. Any quaternionic matrix $ G\in M_{2N}(\mathbb{C}) $ which has $4N^2$ degrees of freedom is similar to an upper-triangular matrix (see \cite{Loring}), called 
{Schur decomposition}. That is
	\begin{equation}\label{eq:Rdef}
	G = U \widetilde{G} U^\dagger\ , 
	\end{equation}
	with a unitary symplectic matrix $ U \in$ USp$(2N)/$U$(1)^N$, and 
	\begin{align}
	\widetilde{G} = 
	\begin{pmatrix}
	\multicolumn{1}{c|}{Z_1}  &  &  \multicolumn{2}{c}{T}  \\
	\cline{1-2}							& \multicolumn{1}{|c|}{Z_2}  	&  & \\
	\cline{2-2}							&								& \ddots & \\
	\cline{5-5}			\multicolumn{2}{c}{0} &							&				&	\multicolumn{1}{|c}{Z_N}
	\end{pmatrix}.
	\label{eq:schur_decomposition}
	\end{align}
	In \eqref{eq:schur_decomposition}, the $ Z_i $ are diagonal $ 2\times2 $-matrices
	\begin{align}
	Z_i = \begin{pmatrix}
	z_i & 0 \\ 0 & \conj{z_i}
	\end{pmatrix},
	\end{align}
	containing the standard complex eigenvalue pairs $ z_i$ and $\conj{z_i} $ 
	of $G$, representing $2N$ degrees of freedom. The strictly upper triangular matrix $ T $ inherits the quaternionic structure of $ G $ because $ U $ is symplectic, i.e. $ T $ consists of $ 2\times 2 $-blocks such that 
	\begin{align}
	\begin{pmatrix}
	T_{i,j+1}	&	T_{i,\conj{j+1}} \\ 
	T_{\conj{i},j+1}		&	T_{\conj{i},\conj{j+1}}
	\end{pmatrix} = 
	\begin{pmatrix}
	a			&	b \\
	-\conj{b}	&	\conj{a}
	\end{pmatrix}, \label{eq:T_block_notation}
	\end{align}
	for some complex numbers $ a $ and $ b $, cf. \eqref{eq:embedding_quaternion_to_complex}, 
for all $i,j=1,2,\ldots,N-1$,  where we use the same notation for the (conjugated) indices as in \eqref{eq:bar-notation}.
These are $2N(N-1)$ degrees of freedom.
The Jacobian of this transformation is well known \cite{Ginibre1965,Mehta}, and we present the result for the joint probability distribution function (jpdf) of the induced Ginibre ensemble \eqref{eq:Ginind}:
\begin{align}\label{eq:jpdf_complete_schur}
			P\left({G}\right) \diff \left[{G}\right] =& C_N^\prime
			\prod_{1\leq i<j\leq N} |z_i-z_j|^2 |z_i - \conj{z_j}|^2 \prod_{i=1}^{N} |z_i-\conj{z_i}|^2 |z_i|^{2\alpha}\\ 
			\nonumber &\times \exp\left[ -2\sigma^{-2}\sum_{k=1}^N|z_k|^2 - \frac{1}{\sigma^2} \Tr(TT^\dagger) \right] \diff[z] \diff[T] \diff[S]\ . 
			\end{align}
Here,  $ C_N^\prime $ is a different normalisation constant, 
$\diff[z]=\prod_{i=1}^N\diff^2z_i$, and	the measure $\diff[T]$ is defined as in \eqref{eq:GinW} over all real and imaginary parts of the $2N(N-1)$ independent matrix elements $T_{i,j+1}$ and $T_{i,\conj{j+1}}$ for $i,j=1,2,\ldots,N-1$, see \eqref{eq:T_block_notation}.
We denote by $\diff[S]$ the induced Haar measure of the coset ${\rm USp}(2N)/[{\rm U}(1)]^{N}$ where the division results from the commutation with the diagonal matrix.
For the quasi-harmonic potential the first term in the exponential in \eqref{eq:jpdf_complete_schur} is replaced by $-\sum_{k=1}^N(2a|z_k|^2-V(z_k)-\conj{V(z_k)})$. 
Note that $T$ drops out from the trace over the analytic function $V(\widetilde{G})$.

It is not surprising that the overlaps will depend on the upper triangular matrix $T$, and in both examples \eqref{eq:Ginell} and \eqref{eq:Ginind} we will be able to perform the integrals over $T$ as they are Gaussian.
Notice that in the joint density \eqref{eq:jpdf_complete_schur} we only integrate over $N$ complex eigenvalues, thus 
fixing  an enumeration of $ N $ non-conjugate eigenvalues $z_{i=1,\ldots,N}$. The matrix $G$ then automatically also has  the complex conjugated eigenvalues $\conj{z}_{i=1,\ldots,N}$. 

The joint density \eqref{eq:jpdf_complete_schur} has some further symmetries. First, it is invariant under complex conjugation, or equivalently under the exchange $z_i\leftrightarrow \conj{z_i}$ for each single $i=1,\ldots,N$, and second,
it is invariant under arbitrary permutations of the $z_i$.
This leads us to the conclusion that  self-overlaps for conjugate eigenvalues $\mathcal{O}_{ii}$ and ${\mathcal{O}}_{\conj{i}\conj{i}}  $, are identical for each $i$,  cf. \cite{GD}, as well as
$\mathcal{O}_{i\conj j}$ and $\conj{\mathcal{O}}_{\conj{i}{j}}  $ for each pair $i,j$. 
Second,  due to the permutation symmetry we may 
focus on the index pairs $ i,j\in \{ 1,\conj{1}, 2, \conj{2} \} $ only. 
In the next section we will explicitly compute these overlaps and verify that both symmetries indeed apply.

Following \cite{CM:2000} we now define the conditional expectation (mean) value of the diagonal and off-diagonal overlap\footnote{Notice the difference in normalisation in $\mathcal{O}_N\left(x_1,x_2\right) $ compared to \cite{CM:2000}.}
	\begin{align} 
	\mathcal{O}_N(x) &:= \frac{1}{N}\left\langle\sum_{l=1}^{N} 
	\delta(x-z_l)\mathcal{O}_{ll} 
	\right\rangle,  \stepcounter{equation}\tag{{\theequation}a} \label{eq:onepoint}  \\
	\mathcal{O}_N\left(x_1,x_2\right) &:= \frac{1}{N^2}\left\langle   
	\sum_{ \substack{ k,l=1;\ k\neq l}}^{N}  
	\delta(x_1-z_k)\delta(x_2-z_l)\mathcal{O}_{kl} \right\rangle ,
	\tag{{\theequation}b} \label{eq:twopoint}\\ 
	\widetilde{\mathcal{O}_N}\left(x_1,x_2\right) &:= \frac{1}{N^2}\left\langle   
	\sum_{ \substack{ k,l=1;\ k\neq l}}^{N}  	
	\delta(x_1-{z_k})\delta(x_2-\conj{z_l})\mathcal{O}_{{k}\conj l}  
\right\rangle ,
\nonumber 
	\end{align}
at arguments $x,x_1,x_2\in\mathbb{C}$, respectively.	
The expectation values are taken over the corresponding ensemble of quaternionic matrices, that is here \eqref{eq:jpdf_complete_schur}. 
As mentioned above we find that the diagonal overlap defined for $\mathcal{O}_{\conj{l}\conj{l}} $ agrees with 	the above, $\mathcal{O}_N(x) = \frac{1}{N}\left\langle\sum_{l=1}^{N} 
	\delta\left(x-\conj{z_l}\right)\mathcal{O}_{\conj{l}\conj{l}} \right\rangle$. Similarly, the remaining off-diagonal overlaps defined for $\mathcal{O}_{\conj{k}\conj{l} } $ and $\mathcal{O}_{\conj k {l}} $ can be obtained by complex conjugation of $	\mathcal{O}_N\left(x_1,x_2\right) $ and 
$\widetilde{\mathcal{O}_N}\left(x_1,x_2\right) $, respectively.

For comparison we also define the spectral 1- and 2-point functions, see e.g. \cite{Kanzieper2001}\footnote{Here, contact terms are absent in the 2-point function.} 
\begin{eqnarray}\label{eq:R1}
\varrho_{1,N}(x)&:=&\frac{1}{N}\left\langle\sum_{l=1}^{N} 
	\delta(x-z_l)\right\rangle, \\	
	\label{eq:R12}
	\varrho_{2,N}(x_1,x_2)&:=&\frac{1}{N^2}\left\langle\sum_{ \substack{ k,l=1;\ k\neq l}}^{N}  
	\delta(x_1-z_l)\delta(x_2-z_k)\right\rangle\ .	
\end{eqnarray}
Here, only delta-functions with respect to the $N$ complex eigenvalues $z_{i=1,\ldots,N}$ are inserted, due to the symmetry of \eqref{eq:jpdf_complete_schur}. Defining furthermore the following density that may be useful when using the technique of Green functions, 
\begin{equation}
\label{eq:Ddef}
D(x_1,x_2):=\frac{1}{N}\left\langle\sum_{ \substack{ k,l=1}}^{N}  
	\delta(x_1-z_l)\delta(x_2-z_k)\  \mathcal{O}_{kl}\right\rangle\ ,
\end{equation}
including also the diagonal sum (notice that we sum only over $N$ indices here), it holds that
\begin{equation}
D(x_1,x_2)=\mathcal{O}_N(x_1)\delta(x_1-x_2)+ N\mathcal{O}_N(x_1,x_2)\ .
\end{equation}
Integrating over $x_2$ and using \eqref{eq:Oid2} we arrive at the following relation to the spectral density:
\begin{equation}
\int \diff^2x_2 D(x_1,x_2)=\frac{1}{N}\left\langle\sum_{l=1}^{N} 
	\delta(x_1-z_l)\right\rangle= \varrho_{1,N}(x_1) \ ,
\end{equation}
exploiting the symmetry of \eqref{eq:jpdf_complete_schur}.
This is all in complete analogy to the complex Ginibre case \cite{CM:2000}.

	 
\sect{Computation of the mean overlaps at finite $N$}
\label{sec:inductive_calculation}
	
In this section we explicitly compute the expectation values of the diagonal and off-diagonal overlaps. We will do this in two steps. First, the expectation value with respect to the upper triangular matrix $T$ from the Schur decomposition \eqref{eq:schur_decomposition} will be taken, which is Gaussian for the quasi-harmonic and induced Ginibre ensemble that we consider here, cf. \eqref{eq:jpdf_complete_schur}. 
Because the integral over the unitary symplectic matrix $U$ decouples, we are left with an average over the complex eigenvalues, which will be computed in the second 	step. 
At this point we will only be able to give explicit results for the rotationally invariant induced Ginibre ensemble. Otherwise our results are still valid for any quasi-harmonic potential.

	\subsection{$T$-average of the diagonal overlaps}\label{sec:diagO}
	
	We start by computing the expectation of $ \mathcal{O}_{11}$.	
For the first step	we can follow the work of Mehlig and Chalker \cite{CM:2000} closely, beginning with the first eigenvectors.
Recalling our notation \eqref{eq:bar-notation},
in a fixed basis corresponding to the Schur normal form \eqref{eq:schur_decomposition}, the normalised first right eigenvector is represented by the first basis vector. This fixes the constant in \eqref{eq:trafo}, and due to the bi-orthogonality condition \eqref{eq:bi_orthogonality} implies that in the first left eigenvector the first component is unity. Its eigenvalue equation \eqref{Lev} and the Schur normal form \eqref{eq:schur_decomposition} require the second component to vanish. In particular,
	\begin{align}
	R_1 &= (1,\; 0,\; \dots\;,\; 0)^T, \\
	L_1^\dagger &= (1,\; 0,\; b_2,\; b_{\conj 2},\; \dots\;,\; b_N,\; b_{\conj N} ),  \label{eq:L_1}
	\end{align}
	are right and left eigenvectors of $ \widetilde G $ corresponding to the eigenvalue $ z_1 $. The components $ b_i,b_{\conj i}\in \mathbb{C} $, for $i=1,\ldots,N$, where we define $ b_1=1 $ and $ b_{\conj 1}=0 $ for later convenience, depend on all other eigenvalues and matrix elements of $T$, 
to be determined recursively below. In addition, we have from \eqref{RRorth} 
that $ R_{\conj{1}} = \hat{\tau}_2 \conj{R_1} $, and since Hermitian adjoints of left eigenvectors of $ \widetilde G $ are right eigenvectors of $ \widetilde G^\dagger $, we conclude $ L_{\conj{1}}^\dagger = \conj{L_1}^\dagger\hat{\tau}_2 $ as well, cf. \eqref{LLorth}. This implies that
	\begin{align}
	R_{\conj{1}} &= \imath(0,\; 1,\; 0,\;\dots\; , \; 0)^T \ ,\label{eq:R_bar_1}\\ 
	L_{\conj{1}}^\dagger &= \imath (0,\; -1,\; \conj{b_{\conj 2}},\; -\conj{b_2},\; \dots\; , \; \conj{b_{\conj N}}, \; -\conj{b_N})\ , \label{eq:L_bar_1} 
	\end{align}
	are  right and left eigenvectors of $ \widetilde{G} $ to $ \conj{z_1} $. 
Consequently the first (diagonal) overlap matrix element 
$ \mathcal{O}_{11}= L_1^\dagger\cdot L_1\  R_1^\dagger \cdot R_1$
can be expressed  as
	\begin{align}\label{eq:O11bb}
	\mathcal{O}_{11} = \sum_{k=1}^{N} \left( |{b_k}|^2 + |{b_{\conj k}}|^2 \right),
	\end{align}	
and we also obtain $\mathcal{O}_{\conj{1}\conj{1}} =\mathcal{O}_{11} $, as previously found in \cite{GD}.
	Because we chose the first right eigenvector $R_1$ to have norm of unity this gives the squared norm of the first left eigenvector $L_1$. 
	
In the following we will take the Gaussian average of \eqref{eq:O11bb}  over the matrix $T$ from \eqref{eq:schur_decomposition}.
	The main idea is to use the eigenvalue equations \eqref{eq:bar-notation} for the left eigenvector, to 
	 obtain recursive relations for the components $ b_i $ and $b_{\conj i}$. 
	Imposing $ L_1^\dagger\widetilde{G}=z_1 L_1^\dagger $ 
and 	$ L_{\conj 1}^\dagger\widetilde{G}=\conj{z_1} L_{\conj 1}^\dagger $ 
	we obtain equations for each pair of columns {with} $ p>1 $:
	\begin{align}
	(T_{1,p},\; T_{1,\conj p}) +(b_p,b_{\conj p}) \begin{pmatrix} z_p & 0 \\ 0 & \conj{z_p}	\end{pmatrix} + \sum_{k=2}^{p-1} (b_k,\; b_{\conj k}) \begin{pmatrix} T_{k, p} & T_{k, \conj p} \\ T_{\conj{k},p} & T_{\conj{k},\conj{p}} \end{pmatrix} = z_1 (b_p,\; b_{\conj p}).
	\end{align}
	Recalling $ b_1 = 1, b_{\conj 1}= 0 $, we can extract the following recursions for $ b_p$ and  $b_{\conj p} $ for $ p>1 $:
	\begin{align*}
	b_p &= \frac{1}{z_1-z_p} \sum_{k=1}^{p-1} \left( b_kT_{k,p} + b_{\conj k} T_{\conj{k},p}   \right), \stepcounter{equation}\tag{{\theequation}a} \label{eq:recursion_for_b}\\
	b_{\conj p} &= \frac{1}{z_1-\conj{z_p}} \sum_{k=1}^{p-1} \left( b_kT_{k,\conj p} + b_{\conj k} T_{\conj{k},\conj p}   \right). \tag{{\theequation}b} \label{eq:recursion_for_b'}
	\end{align*}
	For example, the coefficients up to $ p=3 $ are 
	\begin{align*}
	&b_2 = \frac{T_{12}}{z_1-z_2}, \stepcounter{equation}\tag{{\theequation}a}\label{eq:recursion_for_b_example_1} \\
	&b_{\conj 2} =\frac{T_{1\conj 2}}{z_1-\conj{z_2}}, \tag{{\theequation}b} \label{eq:recursion_for_b'_example}\\
	&b_3 = \frac{T_{13}}{z_1-z_3} + \frac{T_{12} T_{23}}{(z_1-z_2)(z_1-z_3)} + \frac{T_{1\conj 2} T_{\conj{2}3}}{(z_1-\conj{z_2})(z_1-z_3)}, \tag{{\theequation}c}
	\label{T23}\\
	&b_{\conj 3} = \frac{T_{1\conj 3}}{z_1-\conj{z_3}} + \frac{T_{12} T_{2\conj 3}}{(z_1-z_2)(z_1-\conj{z_3})} + \frac{T_{1\conj 2} T_{\conj{23}}}{(z_1-\conj{z_2})(z_1-\conj{z_3})}. \tag{{\theequation}d}\label{T2b3}
	\end{align*}	
	These and all further coefficients are completely determined by $ T $, the eigenvalues $z_i$ and their complex conjugates $\conj{z_i}$ . Note that above we have not yet used that the $2\times 2$ blocks of $T$ represent quaternions and that thus its elements are related as in \eqref{eq:T_block_notation}. Consequently, in the above example \eqref{T23}- \eqref{T2b3} we have that $T_{\conj{23}}=\conj{T_{23}}$ and $T_{\conj{2}3}=-\conj{T_{2\conj 3}}$.

	In order to find the expectation value \eqref{eq:onepoint} of $ \mathcal{O}_{11} $, using \eqref{eq:jpdf_complete_schur} for our ensembles, we evaluate the $T$-integrals first and the more involved eigenvalue integrations in the following Subsection \ref{sec:EvO11}.
We define the $T $-expectation value with respect to the normalised density as
	\begin{align}\label{eq:T-exp}
	\langle\ Q(T)\ \rangle_T:=& \int Q(T)
	\exp[-\sigma^{-2}\Tr (TT^\dag)] \diff[T] \\
	=& C_N^\prime \int Q(T)\ \e^{
	-2\sigma^{-2}\sum_{i,j=1}^{N-1}\left(|T_{i,j+1}|^2+|T_{i,\conj{j+1}}|^2\right)}
\diff[T] \ ,
	 \nonumber
	\end{align}
	with $C_N^\prime=(2/\pi\sigma^2)^{N(N-1)}$. We only sum over a set of independent matrix elements, using \eqref{eq:T_block_notation} which gives a factor of 2 in the exponent.
	This corresponds to a strictly upper triangular, quaternionic matrix $T$ with identically and independently distributed Gaussian entries. 
	Now $ \langle{\mathcal{O}_{11}}\rangle_T $ can be calculated recursively. To begin we define
	\begin{align}
	S_l :=  \left\langle\sum_{p=1}^{l} \left( { |{b_p}|^2 } + { |{b_{\conj p}}|^2 } \right)\right\rangle_T.
	\end{align}
	Obviously, $ S_1 = 1 $ and $ S_N=\langle{\mathcal{O}_{11}}\rangle_T  =\langle{\mathcal{O}_{\conj 1\conj 1}}\rangle_T $ hold. Due to the independent distributions of all 
\begin{equation}
T_{i,j+1}=\conj{ T_{\conj{i},\conj{j+1}} } \quad \mbox{and}\quad T_{i,\conj{j+1}} = -\conj{T_{\conj{i},j+1}} ,
\label{Trel}
\end{equation}
 cf. \eqref{eq:T_block_notation}, the summands in $ S_l $ simplify considerably. Consider for example for $p>1$
	\begin{align}
	|{b_p}|^2 = \frac{1}{|{z_1-z_p}|^2} \sum_{k,l=1}^{p-1} \left( b_kT_{k,p} + b_{\conj k} T_{\conj{k},p}   \right) \conj{\left( b_lT_{l,p} + b_{\conj l}T_{\conj{l},p}   \right)}.
	\end{align}
	Since $ b_k $ and $ T_{k,p} $ are always independent for $ k\leq p-1 $ (see \eqref{eq:recursion_for_b}), the $ T $-average for $ l < k $,
	\begin{align}\label{eq:T-exp-factorise}
	\left\langle{ b_k  T_{k,p} \conj{b_lT_{l,p}} }\right\rangle_T = \left\langle{ T_{k,p} }\right\rangle_T \left\langle{b_k \conj{b_lT_{l,p}}   }\right\rangle_T = 0\ ,
	\end{align}  
	must vanish, as all entries of $ T $ have mean zero. Analogously, the same holds true for $ k<l $,
	\begin{align}
\left\langle	{b_k  T_{k,p} \conj{b_lT_{l,p}}}\right\rangle_T = 0.
	\end{align}
	A similar argument applies to the terms containing products of $ b_{\conj k} $ and $ \conj{b_{\conj l}} $, and in addition to all mixed terms with $ b_k $ and $ \conj{b_{\conj l}} $, or $ b_{\conj k} $ and $ \conj{b_l} $. The result is that only summands featuring $ k=l $ survive when taking the expectation value over $ T $, leading to
	\begin{align*}
	\left\langle{|{b_p}|^2}\right\rangle_T &= \frac{1}{|{z_1-z_p}|^2} \sum_{k=1}^{p-1} \left\langle{ |{b_k}|^2 |{T_{k,p}}|^2 + |{b_{\conj k}}|^2 |{T_{\conj{k},p}}|^2 }\right\rangle_T, \stepcounter{equation}\tag{{\theequation}a}\\
	\left\langle{|{b_{\conj p}}|^2}\right\rangle_T&= \frac{1}{|{z_1-\conj{z_p}}|^2} \sum_{k=1}^{p-1} \left\langle{ |{b_k}|^2 |{T_{k,\conj p}}|^2 + |{b_{\conj k}}|^2 |{T_{\conj{k},\conj{p}}}|^2 }\right\rangle_T \tag{{\theequation}b},
	\end{align*}
	where we used \eqref{Trel}.
	The remaining expectation values factorise completely
for $k\leq p-1$, that is
	\begin{align*}
	\left\langle{ |{b_k}|^2 |{T_{k,p} }|^2}\right\rangle_T &= \left\langle{ |{b_k}|^2 }\right\rangle_T \left\langle{ |{T_{k,p}}|^2 }\right\rangle_T = \frac{\sigma^2}{2}\left\langle{ |{b_k}|^2 }\right\rangle_T,
\stepcounter{equation}\tag{{\theequation}a}\\
		\left\langle{ |{b_{\conj k}}|^2 |{T_{k,\conj{p}} }|^2}\right\rangle_T &= \left\langle{ |{b_{\conj k}}|^2 }\right\rangle_T \left\langle{ |{T_{k,\conj{p}}}|^2 }\right\rangle_T = \frac{\sigma^2}{2}\left\langle{ |{b_{\conj k}}|^2 }\right\rangle_T
	\tag{{\theequation}b}.
	\end{align*}
	Hence  
	for $p>1$
	\begin{align*}
	\left\langle{ |{b_p}|^2 }\right\rangle_T &= \frac{\sigma^2}{2 |{z_1-z_p}|^2 } S_{p-1}, \stepcounter{equation}\tag{{\theequation}a}\\
	\left\langle{ |{b_{\conj p}}|^2 }\right\rangle_T &= \frac{\sigma^2}{2 |{z_1-\conj{z_p}}|^2 } S_{p-1}, \tag{{\theequation}b}
	\end{align*}
	by which we see
	\begin{align}
	S_l = \sum_{p=2}^{l} \left( \frac{\sigma^2}{2 |{z_1-z_p}|^2} + \frac{\sigma^2}{2 |{z_1-\conj{z_p}}|^2} \right)S_{p-1} +1.
	\end{align}
	We can split the sum in two, one part with $ p<l $ and another one having only $ p=l $, to arrive at
	\begin{align}
	S_l = S_{l-1}+\frac{\sigma^2}{2} \left( \frac{1}{|{z_1-z_l}|^2} + \frac{1}{|{z_1-\conj{z_l}}|^2} \right)S_{l-1}.
	\end{align}
	This inductive relation together with the initial condition $ S_1=1 $ yields the final result for the $T$-expectation value of the diagonal overlap
	\begin{align}
	S_N = \left\langle{\mathcal{O}_{11}}\right\rangle_T=\langle{\mathcal{O}_{\conj 1\conj 1}}\rangle_T  = \prod_{l=2}^{N} \left( 1+ \frac{\sigma^2}{2|{z_1-z_l}|^2} + \frac{\sigma^2}{2|{z_1-\conj{z_l}}|^2} \right). \label{eq:o11_T_expvalue}
	\end{align}
	Evidently, it becomes arbitrarily large if $ z_1 $ approaches another eigenvalue of $\widetilde G$ or its complex conjugate. This singularity is removed after taking also the average over the eigenvalues to obtain $ \mathcal{O}_N(z) $, which we show in Subsection \ref{sec:EvO11}.
	
	\subsection{$T$-average of the off-diagonal overlaps}
	
We now determine the overlap $ \mathcal{O}_{12} = L_1^\dagger \cdot L_2\ R_2^\dagger\cdot R_1 $ for which we additionally need the left and right eigenvectors $ L_2 $ and $ R_2 $ to $ z_2 $. We choose the third component of $ R_2 $ as unity. The following components must be zero, since $ R_2 $ is not an eigenvector to any of the other eigenvalues, which occupy the diagonal in \eqref{eq:schur_decomposition}. For a proper upper triangular matrix $\widetilde G$, however, we cannot expect $ R_2 $ to equal a Cartesian unit vector, and so we have  
	\begin{align} 
	R_2 &= (c_1,\; c_{\conj 1},\; 1,\; 0,\; \dots\; , \; 0)^T,\quad 
	R_{\conj{2}} = \imath (-\conj{c_{\conj 1}},\; \conj{c_1},\; 0,\;1,\; 0,\; \dots\; , \; 0)^T, 
	\end{align}
	for some complex $ c_1 $ and $ c_{\conj 1} $ to be determined later. $ L_2 $ is the eigenvector to the eigenvalue $ z_2 $, so its first two components have to vanish, as does the fourth, due to the form of \eqref{eq:schur_decomposition}. Denoting the remaining components by $ d_i, d_{\conj i} \in \mathbb{C}$, $i=3,l\dots,N$, we obtain 
	\begin{align}	
	L_2^\dagger &= (0,\; 0,\; 1,\; 0,\; d_3,\; d_{\conj 3},\; \dots\; , \; d_N,\; d_{\conj N}), \quad
	L_{\conj{2}}^\dagger = \imath(0,\; 0,\; 0,\; -1,\; \conj{d_{\conj 3}},\; -\conj{d_3},\; \dots\; , \; \conj{d_{\conj N}},\; -\conj{d_N}).
	\end{align}
	As for $ z_1 $, we write $ L_2^\dagger \widetilde{G} = z_2 L_2^\dagger $ by pairs of columns with $p>2$:
	\begin{align}
	(T_{2,p}, T_{2,\conj p}) + (d_p,d_{\conj p}) \begin{pmatrix} z_p & 0 \\ 0 & \conj{z_p} \end{pmatrix} + \sum_{k=3}^{p-1} (d_k,d_{\conj k}) \begin{pmatrix} T_{k,p} & T_{k,\conj p} \\ T_{\conj{k},p} & T_{\conj{k},\conj p} \end{pmatrix} = z_2(d_p,d_{\conj p}).
	\end{align}
	Thus, we find relations similar to \eqref{eq:recursion_for_b} and \eqref{eq:recursion_for_b'},
	\begin{align*}
	d_p &= \frac{1}{z_2-z_p} \sum_{k=2}^{p-1} \left( d_kT_{k,p} + d_{\conj k}T_{\conj{k},p}   \right), \stepcounter{equation}\tag{{\theequation}a} \label{eq:recursion_for_d}\\
	d_{\conj p}&= \frac{1}{z_2-\conj{z_p}} \sum_{k=2}^{p-1} \left( d_kT_{k,\conj p} + d_{\conj k}T_{\conj{k},\conj p}   \right), \tag{{\theequation}b} \label{eq:recursion_for_d'}
	\end{align*}
	in which we have set $ d_{\conj 2}=0 $ and $ d_2 = 1. $ The constants $ c_1 $ and $ c_{\conj 1} $ can be found due to the orthogonality of $ R_2 $ with  $ L_1 $ and $ L_{\conj{1}}^\dagger $, viz. $ c_1=-b_2 $ and $ c_{\conj 1}=\conj{b_{\conj 2}}. $ Collecting everything, we obtain 
	\begin{align}
	\mathcal{O}_{12} = \left( b_2 + \sum_{k=3}^{N} \left( b_k\conj{d_k} + b_{\conj k}\conj{d_{\conj k}} \right) \right)\conj{c_1} = -\conj{b_2} \sum_{k=2}^{N}\left( b_k\conj{d_k} + b_{\conj k}\conj{d_{\conj k}} \right). \label{eq:o12-formula}
	\end{align}
It is not difficult to check that $\mathcal{O}_{\conj{1}\conj{2}}=\conj{\mathcal{O}_{12}}$, and for completeness we also give	
		\begin{align}
	\mathcal{O}_{1\conj{2}} = \conj{b_{\conj 2}} \sum_{k=2}^{N}\left( b_k{d_{\conj k}} - b_{\conj k}{d_k} \right)=\conj{\mathcal{O}_{\conj{1}{2}}}. \label{eq:o1b2-formula}
	\end{align}
Evaluating 	$\mathcal{O}_{12} $ from \eqref{eq:o12-formula}, 
as for $ \mathcal{O}_{11} $, the $T$-integrals in the expectation value of $ \mathcal{O}_{12} $ are computed recursively. We set
	\begin{align}
	U_l := \left\langle{ \conj{T_{12}} \sum_{k=2}^{l} \left( b_k \conj{d_k} + b_{\conj k}\conj{d_{\conj k}} \right) }\right\rangle_T \label{eq:U_l_definition}
	\end{align}
for $l\geq2$,	and note that \eqref{eq:recursion_for_b_example_1} implies 
	\begin{align}
	U_2 = \left\langle{ \conj{T_{12}} b_2 }\right\rangle_T = \frac{1}{z_1-z_2}\left\langle{|{T_{12}}|^2 }\right\rangle_T = \frac{\sigma^2}{2(z_1-z_2)}\ . \label{eq:U_l_Anfangswert}
	\end{align}
	Likewise \eqref{eq:o12-formula}, \eqref{eq:recursion_for_b_example_1} and \eqref{eq:U_l_definition} give
	\begin{align}
	U_N = -(\conj{z_1}-\conj{z_2}) \lla{ \mathcal{O}_{12} }\rra_T.
	\end{align}
	In the next step, we consider the difference
	\begin{align}
	U_{l+1} - U_l = \lla{ \conj{T_{12}} \left(b_{l+1} \conj{d_{l+1}} + b_{\conj{l+1}}\conj{d_{\conj{l+1}}} \right) }\rra_T ,
	\end{align}
	and employ \eqref{eq:recursion_for_b}, \eqref{eq:recursion_for_b'}, \eqref{eq:recursion_for_d} and \eqref{eq:recursion_for_d'}, to find for the right hand side
	\begin{align}
	U_{l+1} - U_l =&\lla{  \frac{\conj{T_{12}}}{z_1-z_{l+1}} \sum_{i=1}^{l} \left(  b_iT_{i,l+1} + b_{\conj i}T_{\conj{i},l+1} \right) \frac{1}{\conj{z_2} - \conj{z_{l+1}}}\sum_{j=2}^{l} \left( \conj{d_j T_{j,l+1}} + \conj{d_{\conj j} T_{\conj{j},l+1}} \right)    }\rra_T  
	\\
	& +	 \lla{\frac{\conj{T_{12}}}{z_1-\conj{z_{l+1}}} \sum_{i=1}^{l} \left( b_i T_{i,\conj{l+1}} + b_{\conj i}T_{\conj{i},\conj{l+1}} \right) \frac{1}{\conj{z_2}-z_{l+1}} \sum_{j=2}^{l} \left( \conj{d_j T_{j,\conj{l+1}}} + \conj{d_{\conj j} T_{\conj{j},\conj{l+1}}}  \right) }\rra_T. \nonumber
	\end{align}
	Most of the terms in the expanded products vanish when taking expectation values, because the corresponding matrix elements of $ T $ are independently distributed. More precisely, the recursions \eqref{eq:recursion_for_b}, \eqref{eq:recursion_for_b'}, \eqref{eq:recursion_for_d} and \eqref{eq:recursion_for_d'} for $ b_k,b_{\conj k},d_k,d_{\conj k} $ imply that e.g. $ b_i $ and $ T_{i,l+1} $ are independent if $ i\leq l $. 
This argument leads to the following simplifications, where we use the notation $L\in\{l,\conj{l}\}$:
	\begin{align*}
	\lla{b_i T_{i,L} \conj{d_j T_{j,L}}}\rra_T &= \delta_{ij} \frac{\sigma^2}{2} \lla{b_i\conj{d_j}}\rra_T, \stepcounter{equation}\tag{{\theequation}a} \\
	\lla{b_{\conj i} T_{\conj{i},L} \conj{d_{\conj j} T_{\conj{j},L}}}\rra_T &= \delta_{ij} \frac{\sigma^2}{2}\lla{b_{\conj i}\conj{d_{\conj j}}}\rra_T, \tag{{\theequation}b}\\
	\lla{b_i T_{i,L} \conj{d_{\conj j} T_{\conj{j},L}}}\rra_T &= 0,
	\tag{{\theequation}c}\\
	\lla{b_{\conj i}T_{\conj{i},L} \conj{d_j T_{j,L}}}\rra_T &= 0.
	\tag{{\theequation}d}
	\end{align*}	
	Therefore we arrive at the formula
	\begin{align}
U_{l+1}-U_l \	&= \left(\frac{\sigma^2}{2(z_1-z_{l+1})(\conj{z_2}-\conj{z_{l+1}})} + \frac{\sigma^2}{2( z_1-\conj{z_{l+1}} )( \conj{z_2}-z_{l+1} )} \right) \lla{\conj{T_{12}} \sum_{k=2}^{l} \left( b_k\conj{d_k} + b_{\conj k}\conj{d_{\conj k}} \right)  }\rra_T  \\ \nonumber
	&= \left(\frac{\sigma^2}{2(z_1-z_{l+1})(\conj{z_2}-\conj{z_{l+1}})} + \frac{\sigma^2}{2( z_1-\conj{z_{l+1}} )( \conj{z_2}-z_{l+1} )} \right) U_l.
	\end{align}
	Solving the recursion for $ -U_{N}/(\conj{z_1}-\conj{z_2}) = \lla{\mathcal{O}_{12}}\rra_T  $ yields
	\begin{align}
	\lla{\mathcal{O}_{12}}\rra_T = 	\frac{-\sigma^2}{2 |{z_1-z_2}|^2} \prod_{l=3}^{N} \left(1+ \frac{\sigma^2}{2(z_1-z_{l})(\conj{z_2}-\conj{z_{l}})} + \frac{\sigma^2}{2( z_1-\conj{z_{l}} )( \conj{z_2}-z_{l} )} \right), \label{eq:o12_T_expvalue}
	\end{align}
	due to the initial condition 
	\eqref{eq:U_l_Anfangswert}. 
{As we can see, the $ T $-average of the off-diagonal overlap diverges for $ z_1 $ approaching $ z_2 $ (or vice-versa).}

		Our findings \eqref{eq:o11_T_expvalue} and \eqref{eq:o12_T_expvalue} up to now are remarkable since they solely depend on the variance $ \sigma^2 $ and all the eigenvalues of matrix $ G $, but not on the quasi-harmonic potential. It is therefore clear that these two expressions are symmetric under any change of bases that permutes the blocks $ Z_i $ in \eqref{eq:schur_decomposition}, or that exchanges $ z_i $ and $ \conj{z_i} $ for any $ 1\leq i \leq N $. In other words, replacing $ z_1 $ (and $ z_2 $) by any other eigenvalue $ z_i $ (and $ z_j $) of $ G $ in \eqref{eq:o11_T_expvalue} (and \eqref{eq:o12_T_expvalue}) gives rise to the appropriate non-orthogonality overlap averaged over $ T $. In particular, we have the following expressions
		\begin{align}
		\label{Ob12}
		\lla{\mathcal{O}_{ {1}\conj 2}}\rra_T &= 
	\lla\conj{	{\mathcal{O}_{\conj 1 {2}}}}\rra_T\ = 
		\frac{-\sigma^2}{2 |{{z_1}-\conj{z_2}}|^2} \prod_{l=3}^{N} \left(1		
		+ \frac{\sigma^2}{2( {z_1}-{z_{l}} )( {z_2}-\conj{z_{l}} )} 
		+ \frac{\sigma^2}{2({z_1}-\conj{z_{l}})({z_2}-{z_{l}})} 
		\right), 
\stepcounter{equation}\tag{{\theequation}a} 
\\
\label{Ob1b2}
		\lla{\mathcal{O}_{\conj{1}\conj{2}}}\rra_T &= \lla\conj{ {\mathcal{O}_{12}}}\rra_T \ .
\tag{{\theequation}b} 
		\end{align}
These are the aforementioned relations.

The averages over matrix $T$ that we have performed hold for general harmonic potentials, including both examples \eqref{eq:Ginell} and \eqref{eq:Ginind}. 
This goes beyond the Gaussian ensemble treated in \cite{CM:2000}. We would like to mention that also in the case of a single-channel scattering the overlaps have been expressed in terms of eigenvalues only \cite{FM}. 
To do the average of the complex eigenvalues explicitly in the next subsection we need an isotropic spectrum and therefore restrict ourselves to the induced Ginibre ensemble \eqref{eq:Ginind}.


\subsection{Eigenvalue average of the diagonal overlaps}
\label{sec:EvO11}
Here, we compute the correlation function $\mathcal{O}_N(x)$, cf. \eqref{eq:onepoint}, as the mean of the diagonal overlaps $\lla O_{ii} \rra_T$ with respect to the  jpdf of the induced Ginibre ensemble \eqref{eq:jpdf_complete_schur}
\begin{align}\label{eq:ind_ew_density}
\mathcal{P}_N^{(\alpha)}(Z)
= C_N	\prod_{1\leq i<j\leq N} |z_i-z_j|^2 |z_i - \conj{z_j}|^2 \prod_{i=1}^{N} |z_i-\conj{z_i}|^2 |z_i|^{2\alpha}\exp\left[-2\sigma^{-2}\sum_{k=1}^N|z_k|^2\right] ,
\end{align}	
given that one of the eigenvalues is conditioned to take the value $x\in \mathbb{C}$.
The normalisation constant easily follows from \cite{Mehta}, reading 
\begin{align}
\label{norm}
C_N^{-1}=N! (2\pi)^{N}\left( \frac{\sigma^2}{2}\right)^{N(N+\alpha+1)}\prod_{j=1}^N\Gamma(\alpha+2j)\ .
\end{align}
To begin with, we notice that the jpdf \eqref{eq:ind_ew_density} and thus the  average over the complex eigenvalues denoted by $\langle\quad\rangle_Z$,
\begin{eqnarray}
	&&	\left\langle\, \lla O_{ii} \rra_T \delta(x-z_i)\,\right\rangle_Z:=
\int_{\mathbb{C}^N} 
\mathcal{P}_N^{(\alpha)}(Z)	
	\lla O_{ii} \rra_T \delta(x-z_i) \diff[z] 	
\label{Zvev}	
	\\ \nonumber
 &=&\  C_N	
	\vert x \vert ^{2\alpha} \vert{x-\conj{x}}\vert^2 e^{-\frac{2\vert{x}\vert^2}{\sigma^2}} \int_{\mathbb{C}^{N-1}} \prod_{{l\neq i}}^{N} \left( 1+ \frac{\sigma^2}{2\vert{x-z_l}\vert^2} + \frac{\sigma^2}{2\vert{x-\conj{z_l}}\vert^2} \right) \prod_{{l\neq i}}^N \vert{x-z_l}\vert^2\vert{x-\conj{z_l}}\vert^2 \\ \nonumber
	&&\times \prod_{ \substack{1\leq k < l \leq N \\ k,l\neq i}} \vert{z_l-z_k}\vert^2\vert{z_l-\conj{z_k}}\vert^2\prod_{{l\neq i}}^{N}\vert{z_l-\conj{z_l}}\vert^2 \vert z_l \vert^{2\alpha} \exp\left[  -2\sigma^{-2} \sum_{{l\neq i}}^N \vert{z_l}\vert^2 \right] 
	\prod_{{l\neq i}}^{N} 
	\diff^2 z_l\ ,
\end{eqnarray}
is invariant under relabelling of the eigenvalues, and under the exchange of any $z_l$ with its conjugate $\conj{z_l}$. Consequently, all $N$ summands in \eqref{eq:onepoint} coincide, leading to
\begin{align}\label{eq:onepoint_change_of_variables}
	\mathcal{O}_N(x) =&  \left\langle\, \lla O_{11} \rra_T \delta(x-z_1)\,\right\rangle_Z
\\\nonumber	
	=&\ C_N \vert x \vert ^{2\alpha} \vert{x-\conj{x}}\vert^2 e^{-\frac{2\vert{x}\vert^2}{\sigma^2}} \int_{\mathbb{C}^{N-1}} \prod_{\substack{ l=2}}^{N} \left( 1+ \frac{\sigma^2}{2\vert{x-z_l}\vert^2} + \frac{\sigma^2}{2\vert{x-\conj{z_l}}\vert^2} \right) \prod_{ l=2}^N \vert{x-z_l}\vert^2\vert{x-\conj{z_l}}\vert^2 \\ \nonumber
	&\times \prod_{ 2\leq k < l \leq N } \vert{z_l-z_k}\vert^2\vert{z_l-\conj{z_k}}\vert^2\prod_{l=2}^{N}\vert{z_l-\conj{z_l}}\vert^2 \vert z_l \vert^{2\alpha} \exp\left[  -2\sigma^{-2} \sum_{ l=2}^N \vert{z_l}\vert^2 \right] \prod_{l=2}^N\diff^2 z_l\ .
\end{align}
Moreover, a successive change of variables $ z_l \mapsto \conj{z_l}$ for $l=2,\dots, N$ shows that the terms proportional to $\sigma^2/2$ give the same contribution.
Multiplying the first two products inside the integrand, 
\begin{align}
	\prod_{\substack{ l=2}}^{N} \left( 1+ \frac{\sigma^2}{\vert{x-\conj{z_l}}\vert^2}  \right) \prod_{ l=2}^N \vert{x-z_l}\vert^2\vert{x-\conj{z_l}}\vert^2 = \prod_{\substack{ l=2}}^{N} \left( \vert{x-z_l}\vert^2\vert{x-\conj{z_l}}\vert^2 + {\sigma^2}\vert{x-z_l}\vert^2  \right),
\end{align}
and noting that we can rewrite
\begin{align}
	\prod_{2\leq k < l \leq N} \vert{z_l-z_k}\vert^2\vert{z_l-\conj{z_k}}\vert^2\prod_{l=2}^{N}\vert{z_l-\conj{z_l}}\vert^2 = \det\left[ z_j^{i-1},\, \conj{z_j}^{i-1} \right]_{\substack{1\leq i \leq 2N-2 \\ 2\leq j \leq N}}  \prod_{l=2}^{N}(z_l-\conj{z_l}),	
	\label{Delta2N}
\end{align}
we can employ de Bruijn's formula \cite{deBruijn1955}
\begin{align}
\int_{\mathbb{C}^{N}}  \det\left[ \phi_i(z_j),\, \psi_i({z_j}) \right]_{\substack{1\leq i \leq 2N \\ 1\leq j \leq N}} \diff[z]
=N!\  \text{Pf} \left[ \int_{\mathbb{C}}  (\phi_i(z)\psi_j(z)-\phi_j(z)\psi_i(z))\diff^2z\right]_{1\leq i,j\leq 2N}\ ,
\label{dB}
\end{align}
to arrive at 
\begin{align}
\label{eq:onepoint_de_bruijn} 
	&\mathcal{O}_N(x) =C_N  (N-1)!\ \vert x \vert ^{2\alpha} \vert{x-\conj{x}}\vert^2 \exp\left[-2\sigma^{-2}\vert{x}\vert^2 \right]     \text{Pf} \left[ D_{ij} \right]_{i,j=1}^{2N-2}\ ,
\end{align}
where $D_{ij}$ are the entries of an antisymmetric matrix determined by
\begin{align}
\label{ddef}
	 D_{ij}=\int_{\mathbb{C}} \left( z^{i-1}\conj{z}^{j-1}-z^{j-1}\conj{z}^{i-1} \right) (z-\conj{z}) \vert z \vert^{2\alpha} \left( \vert{x-z}\vert^2\vert{ x-\conj{z} }\vert^2 + {\sigma^2}\vert{x-z}\vert^2  \right) e^{ -2\sigma^{-2}\vert{z}\vert^2} \diff^2z\ . 
\end{align}
Before we embark upon solving this complex integral, we recall  that $ \int_{\mathbb{C}} z^n\conj{z}^m e^{-\frac{2z^2}{\sigma^2}} \diff^2 z$ is only non-zero if $n=m$. Due to the factors $(z-\conj{z})$, $\vert{x-z}\vert^2$ and $\vert{ x-\conj{z} }\vert^2$, the powers in $z^{i-1}\conj{z}^{j-1}$ (and in its conjugate) collect a maximal increment of three, such that $D_{ij}$ is zero whenever $i$ and $j$ differ by more than three. In other words, the first three upper and lower off-diagonals of the matrix with entries $D_{ij}$ are non-zero, in general, whereas all other elements vanish.

The integral involved in $D_{ij} $ can be calculated analytically and is conveniently expressed in terms of $x$, $\alpha$, $\sigma$, $j$ and $k:=j-i=1,2,3$. To this end, we replace $i$ by $j-k$, change to polar coordinates for the integration over the complex plane and expand
\begin{align}
|x-z|^2=	\vert x-re^{\imath\varphi}\vert^2 = \vert x\vert^2 + r^2 -xre^{-\imath\varphi}-\conj{x}re^{\imath\varphi}.
\end{align}
In total, this produces 
\begin{align}
	D_{j-k,j} = &\int_{0}^\infty \int_0^{2\pi} r^{2\alpha+2j-k-1} \left[ e^{\imath\varphi (1-k)} -e^{\imath\varphi(-1-k)} - e^{\imath\varphi(1+k)} + e^{\imath\varphi(-1+k)}\right] \\ \nonumber
	 &\times \Big[ \vert x\vert^4 + r^4 + \left(2\vert x\vert^2 + x^2 + \conj{x}^2\right) r^2 - \left(\vert x \vert^2r +r^3\right) (x+\conj{x})\left( e^{-\imath\varphi} + e^{\imath\varphi}\right) + \vert x\vert^2 r^2 \left( e^{-2\imath\varphi} + e^{2\imath\varphi} \right)  \\ 
	 &\quad\ +{\sigma^2}\left( \vert x\vert^2 +r^2 -xre^{-\imath\varphi}-\conj{x}re^{\imath\varphi}	\right) \Big] e^{-2r^2/\sigma^2}r\diff\varphi\diff r\ . \nonumber 
\end{align}
The $ \varphi $-integration is only non-zero for vanishing exponents, e.g. the integral over $\exp(\imath\varphi(1-k))$ gives rise to $2\pi \delta_{k,1}$. Exploiting this fact yields
\begin{align}
	D_{j-k,j} 
	 =&   4\pi \int_{0}^\infty r^{2\alpha+2j}\Bigg[     \left( \vert x\vert^4 + r^4 + \left(\vert x\vert^2 + x^2 + \conj{x}^2\right) r^2  + {\sigma^2}\left( \vert x\vert^2 +r^2 \right)   \right) \left( r^{-1}\delta_{k,1} - r\delta_{k,-1} \right) \\ \nonumber
	 &\qquad -  \left(\vert x \vert^2 +r^2 +\frac{\sigma^2}{2}     \right) (x+\conj{x}) \left(  r^{-1}  \delta_{k,2}-  r^{3} \delta_{k,-2} \right)  + \vert x\vert^2  \left(  r^{-1}  \delta_{k,3} -  r^{5} \delta_{k,-3} \right)
	 \Bigg] e^{-2r^2/\sigma^2}\diff r\ .
\end{align}
Since $D_{ij}$ is antisymmetric, full information about the entries is contained in the upper triangular part, 
namely for $k=j-i>0$. If we focus on these entries, we can neglect every $\delta_{k,l}$ with negative $l$, and calculate the $r$-integrals according to
\begin{align}
\int_{0}^\infty r^{2b-1} e^{-2r^2/\sigma^2} \diff r = \frac{\Gamma(b)}{2^{b+1}}\sigma^{2b},\quad b>0. \label{eq:gaussian_radial_integral}
\end{align}
Applying these simplifications we find the formula for the upper triangular part with $k>0$:
\begin{align} 
\nonumber
	D_{j-k,j}  =&   2\pi\Gamma(j+\alpha) \left( \frac{\sigma^2}{2} \right)^{j+\alpha} 
	\left\{
	\delta_{k,1}\left[ \vert x\vert^4+ {\sigma^2}
	\vert x \vert^2 + (j+\alpha) \frac{\sigma^2}{2} \left( 
	|x|^2+x^2+\conj{x}^2 +  (j+\alpha+3)\frac{\sigma^2}{2}\right) \right] 
	\right.\\ 
	&
\left.	\qquad\qquad\qquad\qquad\qquad
	-
	\delta_{k,2} 
	(x+\conj{x})\left[ \vert x \vert^2 +(j+\alpha+1) \frac{\sigma^2}{2}\right] 
	+
	\delta_{k,3} 
	\vert x\vert^2\right\},
	\label{dfinal}
\end{align}
which describes the entries of an anti-symmetric matrix that has non-zero elements only on the first three upper and lower off-diagonals.
Together with \eqref{eq:onepoint_de_bruijn} this completes the computation of the diagonal overlap $\mathcal{O}_N(x)$. We recall that the same result holds for the overlap with respect to $\mathcal{O}_{\conj{11}}$.

Note that along the imaginary axis $x=\imath y$, with $y\in\mathbb{R}$, the second upper diagonal in \eqref{dfinal} vanishes. The resulting checkerboard structure of zero and non-zero entries of $D_{ij}$ reduces the Pfaffian in \eqref{eq:onepoint_de_bruijn} to a determinant.


\subsection{Eigenvalue average of the off-diagonal overlaps}

    Let us now focus on the two-point function \eqref{eq:twopoint} using \eqref{eq:o12_T_expvalue}. By definition $ \mathcal{O}_N(x_1,x_2) $ is symmetric under permutation of the eigenvalues $ z_1,\conj{z_1} \dots, z_N, \conj{z_N} $ of $\widetilde G $, cf. Eq.~\eqref{eq:twopoint}. Now we know that $ \langle \mathcal{O}_{ij} \rangle_T $ for $ i\neq j $ is the expression for $ \langle \mathcal{O}_{12} \rangle_T $ with $ z_1 $ replaced by $ z_i $ and $ z_2 $ replaced by $ z_j $, and we also know that $ P_N^{(\alpha)}(Z)$ 
in \eqref{eq:ind_ew_density} 
     is invariant under permutations of the eigenvalues $ z_i $ ($ 1\leq i \leq N $), {hence} 
         \begin{align}
         \label{O12vev}
         	\mathcal{O}_N(x_1,x_2) =\frac{1}{N^2} 
         	N(N-1)\Big\langle &
         	\delta(x_1-z_1)\delta(x_2-z_2)\lla O_{12} \rra_T 
         	\Big\rangle_Z. 
         \end{align}   
    This expression corresponds to the integral 
    \begin{align} 
    \mathcal{O}_N(x_1,x_2) 
    =&-\frac{(N-1)C_N\sigma^2}{2N\vert x_1-x_2 \vert^2} \vert x_1-\conj{x_1} \vert ^2 \vert x_1 \vert^{2\alpha}   \vert x_2-\conj{x_2} \vert ^2  \vert x_2 \vert ^{2\alpha} \vert x_1-x_2 \vert ^2\vert x_1-\conj{x_2} \vert ^2   e^{ -\frac{2\vert x_1 \vert ^2+2\vert x_2 \vert^2}{\sigma^2} }  
\nonumber    
    \\ \nonumber	
    &\times\int_{\mathbb{C}^{N-2}} \prod_{l=3}^N \left( 1+ \frac{\sigma^2}{2(x_1-z_l)(\conj{x_2}-\conj{z_l})}+\frac{\sigma^2}{2(x_1-\conj{z_l})(\conj{x_2}-z_l)} \right)        \det\left[ z_j^{i-1},\, \conj{z_j}^{i-1} \right]_{\substack{1\leq i \leq 2N-4 \\ 3\leq j \leq N}}   \\ 
    &\quad\quad\quad\times 
    \prod_{l=3}^{N} \vert x_1-z_l\vert^2\vert x_1-\conj{z_l} \vert ^2\vert x_2-z_l \vert^2\vert x_2-\conj{z_l} \vert^2 
(z_l-\conj{z_l})  
    \vert z_l\vert^{2\alpha} 
    e^{-\frac{2\vert z_l \vert^2}{\sigma^2} } \diff^2 z_l\ .
    \label{eq:two_point_calculation_eq1}
    \end{align}
    Using the same argument as for \eqref{eq:onepoint_change_of_variables}, we combine the two summands carrying the factor $ \sigma^2$.
    Let us  focus on the 
    integral, 
    which after combining the two products  reads: 
    \begin{align}
    I_{12}:=	\int_{\mathbb{C}^{N-2}} & \prod_{l=3}^N 
    \Big( \vert x_1-z_l\vert^2\vert x_2-z_l \vert ^2+ {\sigma^2}(\conj{x}_1-\conj{z_l})({x_2}-{z_l}) \Big) 
    \vert x_1-\conj{z_l} \vert ^2  \vert x_2-\conj{z_l} \vert ^2 \\ \nonumber
    &\times    
\det\left[ z_j^{i-1},\, \conj{z_j}^{i-1} \right]_{\substack{1\leq i \leq 2N-4 \\ 3\leq j \leq N}}     
   \prod_{l=3}^{N} 
   (z_l-\conj{z_l}) \vert z_l \vert^{2\alpha} e^{-\frac{2 \vert z_l \vert ^2}{\sigma^2} } 
   \diff^2 z_l\ .
    \end{align}  
    De Bruijn's formula \eqref{dB} applied to $ I_{12} $ yields
    \begin{align}
    I_{12} &= (N-2)!\ \text{Pf} \left[  H_{ij}^{12} \right]_{i,j=1}^{2N-4}\ , \quad
     \label{eq:twopoint_de_bruijn}
    \end{align}
    with
    \begin{align}\label{eq:twopoint_de_bruijn_matrix}
    H_{ij}^{12} =& \int_{\mathbb{C}} 
      \left(  \vert x_1-z \vert ^2 \vert x_2-z \vert ^2+{\sigma^2}(\conj{x}_1-\conj{z})(x_2-z)   \right)
     \vert x_1-\conj{z} \vert ^2 \vert x_2-\conj{z} \vert ^2
    \\ \nonumber 
    &\times 
     \left(z^{i-1}\conj{z}^{j-1} -z^{j-1}\conj{z}^{i-1} \right) (z-\conj{z}) \vert z \vert ^{2\alpha}  
   e^{-\frac{2 \vert z \vert ^2}{\sigma^2}}\diff^2 z \ ,
    \end{align}
    the entries of a skew-symmetric matrix.
    We note that the matrix defined by $H_{ij}^{12}$ features five non-zero diagonals above the main diagonal, and in Appendix \ref{appA} we explicitly give these non-vanishing contributions, cf. \eqref{h1} - 
    \eqref{h5}.    
We obtain as a final result for the off-diagonal overlap
        \begin{align} 
   \label{O12final}    
    \mathcal{O}_N(x_1,x_2)=&-\frac{(N-1)!\ C_N\sigma^2}{2N} \vert x_1-\conj{x}_1 \vert^2 \vert x_1\vert^{2\alpha} \vert x_2-\conj{x_2} \vert ^2 \vert x_2 \vert ^{2\alpha}  
    \vert x_1-\conj{x_2} \vert^2 
    \\ 
   &\times \exp\left[ -\frac{2}{\sigma^2}\left(\vert x_1 \vert ^2+\vert x_2 \vert ^2\right) \right] 
   \text{Pf} \left[  H_{ij}^{12} \right]_{i,j=1}^{2N-4} .
   \nonumber  
    \end{align}
 In the same way we can determine the overlap $\widetilde{\mathcal{O}_N}(x_1,x_2)$ by defining a matrix $H_{ij}^{1\conj 2}$, obtained from \eqref{eq:twopoint_de_bruijn_matrix} by interchanging $x_2$ and  $\conj{x_2}$, cf. 
 \eqref{eq:o11_T_expvalue} and \eqref{Ob12}.
The  matrix $H_{ij}^{1\conj 2}$ follows in the same way from Appendix \ref{appA}, and the remaining two off-diagonal overlaps with respect to $\mathcal{O}_{\conj{1}2}$ and $\mathcal{O}_{\conj{12}}$ then follow from complex conjugation.

\sect{The large-{$N$} limit}\label{largeN}

In this section we will analyse the limiting overlaps $ \mathcal{O}_N(x) $ and
$ \mathcal{O}_N(x_1,x_2) $ 
 when $ N \to \infty$.  Because of the extra repulsion of  complex eigenvalue pairs from the real axis in the quaternionic Ginibre ensemble, cf. \eqref{eq:jpdf_complete_schur}, we have to distinguish three regions, the vicinity of the real line, the bulk and the edge of the spectrum. 
Our derivation will be heuristic, for some rigorous results for the diagonal overlap conditioned to $z_1=0$ see \cite{GD}.

The macroscopic density of the quaternionic Ginibre ensemble is known to converge  to the circular law, 
see e.g. \cite[Cor.~2.2]{BenaychGeorges2011},
	\begin{align}
	\varrho_1(x) :=\lim_{N\to\infty} \varrho_{1,N}(z)= 
	\frac{1}{\pi}
	\eins_{|{x}|<1} \ ,
	\label{eq:circular_law}
	\end{align}
where we have chosen $\sigma^2=1/N$ and introduced the standard notation for the characteristic function $\eins_A$ to be non-vanishing only if $A$ is true.
It is known that also in the induced Ginibre ensemble the macroscopic density approaches the circular law, as long as the number of zero-modes $\alpha$ is fixed, cf. \cite{Fischman}. While the density is still constant on 
a different domain for the elliptic Ginibre ensemble (ellipse) and the induced Ginibre ensemble at $\alpha \sim N$ (annulus), the calculations below become more complicated for these ensembles. We expect though that our result remains true in the bulk in these cases as well.

The first subsection is devoted to the bulk of the spectrum, where we  address the macroscopic rather than the local, microscopic behaviour of the overlaps. Here, the results turn out to agree with that of the complex Ginibre ensemble and are thus universal. 
In the second part we address the local behaviour in the vicinity of the origin which is specific to this symmetry class and thus differs from the complex Ginibre ensemble.

	\subsection{Macroscopic bulk limit of the overlaps}
	\label{bulk}
	Here, we will choose the arguments $ x, x_1 $ and $ x_2 $ of the overlaps to be in the bulk of the spectrum. Because the local, microscopic scale of correlations is $O(1/\sqrt{N})$, we choose these arguments to be inside the unit disc with a distance to the boundary (edge) of support and to the real axis of larger order than $\varepsilon= 1/\sqrt{N} $. We will also choose the arguments of the off-diagonal overlap to be at macroscopic distance, $|x_1-x_2| $ to be of larger order than $\varepsilon$.
It is well known that in the macroscopic large-$N$ limit expectation values factorise to leading order. For the above choice of arguments the macroscopic spectral two-point correlation function \eqref{eq:R12} 
thus reads, cf. \cite[Sec.~III.A.3]{Guhr1998},
	\begin{align}
	\varrho_2(x_1,x_2) :=\lim_{N\to\infty}\varrho_{2,N}(x_1,x_2)
	= \varrho_1(x_1)\varrho_1(x_2) = 
\frac{1}{\pi^2}\eins_{|{x_1}|<1}\eins_{|{x_2}|<1}\eins_{x_1\neq x_2}. 
	\label{eq:circular_law_two_evals}
	\end{align}

We begin with the analysis of the diagonal overlap
$ \mathcal{O}_N(x) $. In view of \eqref{eq:onepoint_change_of_variables} we  have for large $ N $
	\begin{align}\label{eq:asymptotics_one-point_expansion}
	\mathcal{O}_N(x) &\approx\left\langle\,\delta(x-z_1)\,\right\rangle_Z \left\langle\, \lla O_{11} \rra_T \,\right\rangle_{Z;z_1=x}\\
	\nonumber
	&= \varrho_1(x)
	\left\langle \prod_{k=2}^{N}\left( 1+\frac{1}{2N|{x-z_k}|^2} + \frac{1}{2N|{x-\conj{z_k}}|^2} \right) \right\rangle_{Z;z_1=x}.
	\end{align}
Here, we also used that due to permutation invariance under the expectation value, the density can be written as $\varrho_{1,N}(z)=\langle \delta(z-z_j)\rangle$, for any $j=1,\ldots,N$.	The density and the average 
$\langle O_{11}\rangle_{Z;z_1=x}$ are over $N-1$ eigenvalues, conditioned to $z_1=x$. 	Rather than computing the large-$N$ average of the product, which we denote by $ \Omega_1 $ we consider its logarithm. It  satisfies
	\begin{align}
	\label{eq:Omega1_expansion} 
\ln \Omega_1:=	
\sum_{k=2}^{N}\ln\left( 1+\frac{1}{2N|{x-z_k}|^2} + \frac{1}{2N|{x-\conj{z_k}}|^2} \right) 
\approx\frac{1}{2N}\sum_{k=2}^{N}\left( \frac{1}{|{x-z_k}|^2}+\frac{1}{|{x-\conj{z_k}}|^2} \right).
	\end{align}
In the following we will assume that we can interchange the exponential and the expectation value at large-$N$, 
\begin{equation}
\langle\ \langle O_{11}\rangle_T\ \rangle_{Z;z_1=x}
=\big\langle\exp[\ln \Omega_1]\big\rangle_{{Z;z_1=x}}
\approx\exp\left[\big\langle\ln \Omega_1\big\rangle_{Z;z_1=x}\right]\ .
\label{Ln-exchange}
\end{equation}	
Due to the average we can combine the two terms in the sum 
\eqref{eq:Omega1_expansion}
to obtain	
	\begin{equation}
	\label{eq:Omega1_expansion2} 
\left\langle\ln \Omega_1\right\rangle_{Z;z_1=x}\approx
\int_{\mathbb{C}} \frac{1}{|{x-z}|^2} \varrho_{1,N-1}(z)\diff^2z
\approx \frac{1}{\pi}
\int_{\mathbb{C}} \eins_{|{z}|<1}\eins_{|{z-x}|>\varepsilon} \frac{\diff^2z}{|{x-z}|^2}=:I(\varepsilon)\ ,
	\end{equation}
with the density \eqref{eq:R1} of $N-1$ variables $z_2,\ldots,z_N$. In the large-$N$ limit the density will converge (under the integral even without taking the average) to the circular law \eqref{eq:circular_law}. In the last step we have regularised the resulting integral by cutting out an open disc $  \{z\in\mathbb{C} \mid |{x-z}|<\varepsilon \} $ around the singularity, because the local scale is of order $O(1/\sqrt{N})$ and our choice\footnote{In \cite{CM:2000} the regularisation is done by removing a small fraction of eigenvalues closest to $x$, without specifying the order in $N$.}  of  $\varepsilon$. 
A rigorous argument is provided in \cite[e.g. Thm. 2.6]{Bourgade2018} for  complex Ginibre matrices, which should be adaptable to our case. 

Our next task is to compute the integral $I(\varepsilon)$ on the right hand side of \eqref{eq:Omega1_expansion2}. We change variables to $ z^\prime=z-x=re^{i\vartheta} $ and then apply a polar decomposition, resulting into
	\begin{align}
\label{I-def}	
I(\varepsilon)=	\frac{1}{\pi} \int_0^{2\pi} \int_\varepsilon^2 \eins_{|{x}|^2+r^2+2r|{x}|\cos(\vartheta)<1} \frac{\diff r\diff\vartheta}{r}\ .
	\end{align}
	Clearly it holds that $|z|,|x|<1$, due to the circular law, and thus $r<2$.
The roots $r_\pm$ of the remaining indicator function
	\begin{align}
f(r):=	1-r^2-2r|{x}|\cos(\vartheta)-|{x}|^2=(r_+(\vartheta)-r)(r- r_-(\vartheta)) \label{eq:asymptotics_integral_indicator_zeroes}
	\end{align}
	 are given by 
	\begin{align}
	r_{\pm}(\vartheta) 
	= -|{x}|\cos(\vartheta)\pm\sqrt{1-|{x}|^2\sin^2(\vartheta)}.
	\end{align}
In order to determine the integration domain in \eqref{I-def} in $r$ we observe that first, $f(r=0)=1-|x|^2>0$ is positive, due to $x$ being in the bulk. 
Furthermore, for $ \vartheta \in [0,\pi/2] $ the continuous function $r_-(\vartheta)<0$ is clearly negative. To show that it remains negative for all angles  we observe that the equation $ r_-(\vartheta_0)=0 $ implies 
	\begin{align}
	|{x}|^2\cos^2(\vartheta_0) = 1-|{x}|^2\sin^2(\vartheta_0) \quad \Leftrightarrow
	\quad 	|{x}|^2 = 1\ ,
	\end{align}
	which is not possible. From the fact that $f(r)$ is a parabola and continuous we conclude that $r_+(\vartheta)>0$ for all angles, restricting the integration domain in \eqref{I-def} to $[\varepsilon ,r_+(\vartheta)]$. We can thus do the radial integral, and use the periodicity of sine and cosine after splitting the integration domain, to obtain
	 \begin{eqnarray}
I(\varepsilon)&=&	
\frac{1}{\pi} \int_0^{2\pi}\ln\left[-|x|\cos(\vartheta)+\sqrt{1-|x|^2\sin^2(\vartheta)}
\right]-\ln(\varepsilon)\,\diff\vartheta
\nonumber\\
&=&\frac{1}{\pi} \int_0^{\pi}\ln\left[-|x|\cos(\vartheta)+\sqrt{1-|x|^2\sin^2(\vartheta)}\right]+\ln\left[|{x}|\cos(\vartheta)+\sqrt{1-|{x}|^2\sin^2(\vartheta)}\right]\diff\vartheta
-2\ln(\varepsilon)
\nonumber\\
&=& \frac{1}{\pi} \int_0^{\pi}\ln\left[{1-|{x}|^2\sin^2(\vartheta)}-|{x}|^2\cos^2(\vartheta)\right]\diff\vartheta
-\ln(\varepsilon^2)\ =\ \ln\left[\frac{1-|x|^2}{\varepsilon^2}\right]\ .
\label{I-result}
	 \end{eqnarray}
Combining \eqref{eq:asymptotics_one-point_expansion} and \eqref{Ln-exchange} we obtain as a final answer from  \eqref{I-result} 
	\begin{align} \label{eq:onepoint_limit}
	\mathcal{O}_N(x) &\approx \frac{N}{\pi}\eins_{|{x}|<1}({1-|{x}|^2})\ ,
	\end{align}	
	recalling that $\varepsilon= 1/\sqrt{N} $.
It holds for the one-parameter family of quaternionic induced Ginibre ensemble with fixed $\alpha$. 	The linear dependence on $N$ indicates the strong sensitivity of the eigenvalues under perturbations.
The result \eqref{eq:onepoint_limit} agrees with the one in \cite{CM:2000} for the complex Ginibre ensemble, and as argued in \cite{FT} can also be expected for the real Ginibre ensemble.  It is thus universal in the sense that it holds for all three Ginibre ensembles sharing the circular law for the global density, and one can expect the same to hold for Wigner ensembles in these three symmetry classes as well. 
However, 
 when moving to products of $m$ complex Ginibre ensembles the circular law is modified to $\frac{1}{m\pi}|x|^{\frac2m-2}$ on the unit disc. The resulting overlap is then multiplied by this density, with the quadratic power in \eqref{eq:onepoint_limit} replaced by $2\to2/m$ \cite{PZ}. 


We go over to the analysis of the off-diagonal overlaps, starting with 	
$ \mathcal{O}_N(x_1,x_2) $. Recalling \eqref{O12vev}, the same factorisation argument as in \eqref{eq:asymptotics_one-point_expansion}
leads to
	\begin{align}
	\label{O12factor}
\mathcal{O}_N(x_1,x_2) &\approx 
\left\langle\,\delta(x_1-z_1)\,\right\rangle_Z \left\langle\,\delta(x_2-z_2)\,\right\rangle_Z\eins_{x_1\neq x_2}
\left\langle\, \lla O_{12} \rra_T \,\right\rangle_{Z;z_{1,2}=x_{1,2}}\\
	\nonumber
	&=  -\frac{ \varrho_1(x_2)\varrho_1(x_2)\eins_{x_1\neq x_2}}{|{x_1-x_2}|^2}
	\nonumber\\
	&\quad\times
	\left\langle 
	 \prod_{l=3}^{N} \left(1+ \frac{1}{2N(x_1-z_{l})(\conj{x_2}-\conj{z_{l}})} + \frac{1}{2N( x_1-\conj{z_{l}} )( \conj{x_2}-z_{l} )} \right)
	\right\rangle_{\!Z;z_{1,2}=x_{1,2}}\!\!\!\!.
	\end{align}
	We proceed as for the diagonal overlap, assuming the following approximation to hold for large-$N$:
	\begin{equation}
	\langle\ \langle O_{12}\rangle_T\ \rangle_{Z;z_{1,2}=x_{1,2}}
=:\big\langle\exp[\ln \Omega_2]\big\rangle_{{Z;z_{1,2}=x_{1,2}}}
\approx\exp\left[\big\langle\ln \Omega_2\big\rangle_{Z;z_{1,2}=x_{1,2}}\right]\ .
\label{Ln-exchange2}
	\end{equation}
We evaluate the logarithm of the product denoted by $\Omega_2$, combining the two terms in the sum to one after averaging. We obtain
	\begin{equation}
	\label{eq:Omega2_expansion2} 
\left\langle\ln \Omega_2\right\rangle_{Z;z_{1,2}=x_{1,2}}
\approx \frac{1}{\pi}
\int_{\mathbb{C}} \eins_{|{z}|<1}\frac{\diff^2z}{(x_1-z)(\conj{x_2}-\conj{z})}=:J(x_1,\conj{x_2})\ .
	\end{equation}	
Here, no regularisation is needed as the two simple poles are integrable in the complex plane. If we repeat the same steps for $\widetilde{\mathcal{O}_N}(x_1,x_2)$, from \eqref{Ob12} we arrive at the same integral, with the replacement $\conj{x_2}\to x_2$. Therefore, in computing the integral  $J(x_1,\conj{x_2})$ we obtain the asymptotic result for all off-diagonal overlaps. In polar coordinates it reads
	\begin{align}
	\label{J-def}
	J(x_1,\conj{x_2})
	= \frac{1}{\pi}\int_{0}^{1} \int_{0}^{2\pi} \frac{\diff \varphi\, r \diff r}{(x_1-re^{i\varphi})(\conj{x_2}-re^{-i\varphi})}=: \frac{1}{\pi}\int_{0}^{1} K(r)r \diff r\ .
	\end{align}
	Let us consider the angular integral $K(r)$ for $0<r<1$:
	\begin{align}
K(r) = \frac{1}{(-r\conj{x_2 })}\int_{0}^{2\pi} \frac{ e^{i\varphi}\diff \varphi }{ (\frac{x_1}{r} - e^{i\varphi}) (\frac{r}{\conj{x_2}} -e^{i\varphi}  ) }. 
	\end{align}
	If we understand this as a contour integral on the unit circle $\gamma$, the integral produces residues, depending on the ratios of the moduli $ |{x_1}| $ and $ |{x_2}| $ to $ r$:
	\begin{align}
	K(r)=& \frac{1}{(-ir\conj{x_2})} \oint_{\gamma} \frac{ \diff \zeta }{ (\frac{x_1}{r} - \zeta) (\frac{r}{\conj{x_2}} - \zeta)  } 
= \frac{\left(\frac{r}{\conj{x_2}}-\frac{x_1}{r}\right)^{-1}}{(-ir\conj{x_2})} \oint_{\gamma} 
\frac{1 }{ (\frac{x_1}{r} - \zeta)} -
\frac{1 }{ (\frac{r}{\conj{x_2}} - \zeta)  } 
\diff \zeta 
	\\ \nonumber
	=& \frac{2\pi }{ r^2 - x_1\conj{x_2}  }   \left(  \eins_{|{x_1}|<r} - \eins_{r<|{x_2}|}  \right).
	\end{align}	
Reinserting into \eqref{J-def}, the radial integral becomes  elementary 
	\begin{align}
J(x_1,\conj{x_2})=& \int_{|{x_1}|}^1 \frac{2r\diff r}{ r^2 - x_1\conj{x_2} } - \int_{0}^{|{x_2}|} \frac{ 2r\diff r}{ r^2 - x_1\conj{x_2}  }
\nonumber\\
	=& \ln\left( \frac{ 1 - x_1\conj{x_2} }{ |{x_1}|^2 - x_1\conj{x_2} } \right) - \ln\left( \frac{ |{x_2}|^2 - x_1\conj{x_2} }{ -x_1\conj{x_2}} \right)  \nonumber\\
	=& \ln\left(
\frac{ 1-x_1\conj{x_2} }{ |{x_1-x_2}|^2 }	\right).
	\end{align}
Putting together \eqref{Ln-exchange2} and \eqref{eq:Omega2_expansion2} we thus arrive at our final result 
	\begin{align}
	\mathcal{O}_N(x_1,x_2) &\approx -\,\frac{1-x_1\conj{x_2}}{\pi^2 |{x_1-x_2}|^4}
\eins_{|{x_1}|<1}\eins_{|{x_2}|<1}	\eins_{x_1\neq x_2}\ .
\label{macroO12final}
	\end{align}
	This, too, agrees with \cite{CM:2000} and is thus universal. The remaining off-diagonal overlap
	$\widetilde{\mathcal{O}_N}(x_1,x_2) $ is obtained by exchanging $\conj{x}_2\leftrightarrow x_2$,  and the two expressions for the overlaps of 
$\mathcal{O}_{\conj{1}\conj{2} } $ and $\mathcal{O}_{\conj 1 {2}} $ follow from complex conjugation. The algebraic decay of the overlaps is in contrast to the exponential decay of the complex eigenvalue correlation functions of the complex and quaternionic Ginibre ensembles, which also agree
in the 
local 
bulk scaling limit \cite{AKMP}.

In \cite{MANT} a list of results for the off-diagonal overlap in various ensembles with complex matrix elements is given, like the induced Ginibre ensemble when $\alpha\sim N$, the truncated unitary, spherical and product ensemble of two Ginibre matrices. In all cases the numerator of \eqref{macroO12final} gets modified, whereas the quartic repulsion is unchanged. This observation goes in line with the observation that in all these examples also the global density differs from the circular law in our case.

\subsection{Microscopic origin limit of the diagonal overlap}\label{origin}

In this subsection we analyse the behaviour of $\mathcal{O}_N(x)$ in the vicinity of the origin. It is clear that the number of zero eigenvalues $2\alpha$ should play a role here. We expect that away from the origin along the real axis similar results hold when setting $\alpha=0$. This is because for the quaternionic Ginibre ensemble without zero modes the origin is representative for all points along the real axis.

Looking at the result for finite $N$, \eqref{eq:onepoint_de_bruijn} together with \eqref{dfinal}, we see that $\mathcal{O}_N(x)$ seems to vainish quadratically $\sim|x-\conj x|^2$ along the real line. On the other hand, if we naively continue the large-$N$ expression \eqref{eq:onepoint_limit} from the bulk (which was derived excluding real $x$) we arrive at  $\lim_{x\to0}\mathcal{O}_N(x)=N/\pi$. 
Therefore, there has to be a transition region that resolves this contradiction. 
To find this regime 
we pursue a more sophisticated method. Writing $ B(\varepsilon) $ for the open ball with radius $ \varepsilon $ and center $ 0 $, we recall Lebesgue's differentiation theorem, which states 
	\begin{align}
	\lim\limits_{\varepsilon\to 0} \langle {f \mid x\in B(\varepsilon)}\rangle = \lim\limits_{\varepsilon\to 0} \frac{\langle{f\eins_B(\varepsilon)}\rangle}{\langle{\eins_B(\varepsilon)}\rangle} = f(0)\ ,
	\end{align}
for $f$ conditioned to take points in $B(\varepsilon)$ on the left hand side. This holds
	given that $ f $ is continuous at $ 0 $. 
We are thus lead to consider the following ratio which we take at finite $N$,  and then take the limit $x\to0$. Inserting \eqref{eq:onepoint_de_bruijn} we can write
	\begin{align}
\lim\limits_{x\to 0} \frac{\mathcal{O}_N(x)}{\varrho_{1,N}(x)}
&=\lim\limits_{x\to 0} \frac{C_N  (N-1)!\ \vert x \vert ^{2\alpha} \vert{x-\conj{x}}\vert^2 \exp\left[-2\sigma^{-2}\vert{x}\vert^2 \right]     \text{Pf} \left[ D_{ij} \right]_{i,j=1}^{2N-2}}{\langle\delta(x-z_1)\rangle_Z}
\nonumber\\
&=\frac{\text{Pf} \left[ 
h_j(j+\alpha+3)\delta_{j-i,1}
-
h_i (i+\alpha+3)
\delta_{i-j,1} \right]_{i,j=1}^{2N-2}}{\text{Pf} \left[ 
h_j(j+\alpha+1) \delta_{j-i,1}
-
h_i(i+\alpha+1) 
\delta_{i-j,1}
 \right]_{i,j=1}^{2N-2}}\ ,
	\end{align}
with 
\begin{equation}
h_j:=2\pi\Gamma(j+\alpha+1) \left( \frac{\sigma^2}{2} \right)^{j+\alpha+2}.
\label{hjdef}
\end{equation}		
Here, we used that the pre-factors of the Pfaffian determinants cancel, and that for the numerator we can simply use \eqref{dfinal} for the upper (and lower) diagonal parts, reinstate $k=j-1$ and set $x=0$. The remaining two upper (and lower) diagonals vanish in this limit $x=0$.
The computation for the denominator $\langle\delta(x-z_1)\rangle_Z$ follows exactly along the lines of Subsection \ref{sec:EvO11} and also simplifies greatly when setting $x=0$. In a last step we may use that the Pfaffian of such a tridiagonal matrix can be easily computed, see e.g. \cite{Mehta},
\begin{equation}
\text{Pf} \left[ h_{j-1}\delta_{j-1,i}-h_{i-1}\delta_{i,j+1}) \right]_{i,j=1}^{2N} =\prod_{j=1}^Nh_{2j-1}\ .
\label{Pfaff}
\end{equation}
This leads us to a telescopic product where most factors cancel, and we arrive at 
\begin{equation}
	\mathcal{O}_N(0)\sim  \frac{1}{\pi}\lim\limits_{x\to 0} \frac{\mathcal{O}_N(x)}{\varrho_{1,N}(x)}
	=  \frac{1}{\pi}
	\prod_{j=1}^{N-1}\frac{(2j+\alpha+3)}{(2j+\alpha+1)}=\frac{2N+\alpha+1}{\pi(3+\alpha)} \ ,
\label{Ozero}
\end{equation}
for the local diagonal overlap at the origin.
We see that for $\alpha=0$ it is close to our above extrapolation based on the macroscopic overlap in the bulk. When the corrections from the $2\alpha$ exact zero eigenvalues become of the order $\alpha=O(N)$, the result \eqref{Ozero} is of order $O(1)$. In that case the circular law gets modified to become a ring \cite{Fischman}, as the zero modes push out the other eigenvalues. Thus 
only very few eigenvalues remain close to the origin, with their number decreasing exponentially in $N$.


\sect{Conclusions}\label{conclusion}

In the present work we have set up the computation of eigenvector correlations in Ginibre ensembles with quaternionic matrix elements, complementing very recent parallel work by Dubach.
Following the ideas of Chalker and Mehlig, we have been able to express the diagonal and off-diagonal overlaps of left and right eigenvectors solely in terms of expectation values of complex eigenvalues, which come in complex conjugate pairs  in this symmetry class. This holds for a general class of ensembles with harmonic potentials, including the elliptic Ginibre ensemble. The different combinations of overlaps between eigenvectors of complex eigenvalues and complex  conjugated ones are all closely related by symmetry. 
In the particular case of the induced quaternionic Ginibre ensemble we could compute the diagonal and off-diagonal overlap explicitly for finite matrix size $N$ and an arbitrary number of zero eigenvalues. These expressions are given by Pfaffian determinants of banded matrices with three respectively five non-vanishing upper diagonals. 

In the large-$N$ limit we computed both diagonal and off-diagonal overlaps in the bulk of the spectrum at macroscopic distance of the eigenvalues. We found agreement with the results for the real and complex Ginibre ensembles. This implies that also for the eigenvector statistics there is no possibility to distinguish these three symmetry classes in the bulk of the spectrum, as was previously found for the eigenvalue spectrum. This has important implications when comparing to the eigenvector statistics in applications. 
In the vicinity of the origin,  which is most likely to be specific for this symmetry class, we investigated the diagonal overlap only.

Several questions remain open for future applications. An integrable Pfaffian structure at finite $N$ is yet to be uncovered for this symmetry class, as could be expected in analogy to the complex ensemble. Such a finding would enable us to take local bulk, origin and edge scaling limits, which are expected to be much more universal than the global statistics. Although global results exist for the elliptic Ginibre ensembles, it would be very interesting to take a weak non-Hermiticity limit for eigenvector statistics, in oder to see the emergence of the independence know in Hermitian eigenvector statistics. In principle the setup we have provided allows to approach this regime.

\section*{Acknowledgments}
Support by the Wallenberg foundation (G.A.), the German research council
DFG through grant  the CRC1283 ``Taming uncertainty and profiting from randomness and low regularity in analysis, stochastics and their applications'' (G.A. and M.K.), as well as by the Studienstiftung des Deutschen Volkes {and EPSRC through EP/L015854/1 Centre for Doctoral Training CANES} (Y.-P.F.) is thankfully acknowledged.
We thank Guillaume Dubach and Dmitry Savin for discussions and correspondence.

\begin{appendix}
\sect{Appendix}\label{appA}

In this appendix we state the results 
for
 the non-vanishing upper  diagonal elements of the antisymmetric matrix $H_{ij}^{12}$ from \eqref{eq:twopoint_de_bruijn_matrix},  appearing in the final expression for the off-diagonal overlap $\mathcal{O}_N(x_1,x_2)$ in \eqref{O12final}.  We repeat its definition for convenience:
    \begin{align}\label{eq:h12}
    H_{ij}^{12} =& \int_{\mathbb{C}} 
      \left(  \vert x_1-z \vert ^2 \vert x_2-z \vert ^2+\frac{\sigma^2}{2}(\conj{x_1}-\conj{z})(x_2-z)   \right)
     \vert x_1-\conj{z} \vert ^2 \vert x_2-\conj{z} \vert ^2
    \\ \nonumber 
    &\quad\times 
     \left(z^{i-1}\conj{z}^{j-1} -z^{j-1}\conj{z}^{i-1} \right) (z-\conj{z}) \vert z \vert ^{2\alpha}  
   e^{-\frac{2 \vert z \vert ^2}{\sigma^2}}\diff^2 z \ .
    \end{align}
The evaluation of $ H_{ij}^{12} $ does not differ technically from the integrations done for \eqref{ddef} for the diagonal overlap, but it is much more laborious. 
For that reason we only give the final answer. 
Again, we define $k:=j-i$, enumerating off-diagonals with respect to the $j$-th column, thus eliminating the row index $i$, and we use polar coordinates  $z=re^{\imath \varphi}$.  
Multiplying out all factors, doing the angular integration that projects onto the non-vanishing five upper (and lower) diagonals, and applying \eqref{eq:gaussian_radial_integral} we arrive at the following expressions.
 First, for $k=1$ we find
            \begin{align}\nonumber
        	H_{j-1,j}^{12} =& 2\pi\Gamma(j+\alpha)\left( \frac{\sigma^{2}}{2} \right)^{j+\alpha} \\ \nonumber
&\times        	\Bigg\{  |x_1|^4|x_2|^4+(j+\alpha) \frac{\sigma^2}{2}\Big[ 
2|x_1|^2|x_2|^2|x_1+x_2|^2    +|x_1|^4(x_2^2+\conj{x_2}^2-|x_2|^2)	
\\ \nonumber
&\qquad\qquad\qquad\qquad\qquad    +|x_2|^4(x_1^2+\conj{x_1}^2-|x_1|^2)  +(x_1x_2+\conj{x_1}\conj{x_2}-x_1\conj{x_2}-\conj{x_1}x_2)|x_1|^2|x_2|^2
        	\Big] 
  \\ \nonumber
  &+(j+\alpha) (j+\alpha+1)\frac{\sigma^4}{4}\Big[
|x_1+x_2|^4+|x_1|^2(2x_2^2+2\conj{x_2}^2-|x_2|^2)  +|x_2|^2(2x_1^2+2\conj{x_1}^2-|x_1|^2)  
\\ \nonumber  
  &+(x_1x_2+\conj{x_1}\conj{x_2})(2|x_1|^2+2|x_2|^2-|x_1+x_2|^2)
  -(x_1\conj{x_2}+\conj{x_1}x_2)(|x_1|^2+|x_2|^2)
  +x_1^2x_2^2+\conj{x_1}^2\conj{x_2}^2
  \Big] 
     \\ \nonumber
        	&\ + (j+\alpha) (j+\alpha+1)(j+\alpha+2)\frac{\sigma^6}{8}\Big[
  |x_1+x_2|^2+x_2^2+\conj{x_2}^2+x_1^2+\conj{x_1}^2
  +x_1x_2+\conj{x_1}\conj{x_2}
        	\Big]
        	    \\ \nonumber
        	&\ + (j+\alpha) (j+\alpha+1)(j+\alpha+2)(j+\alpha+3)\frac{\sigma^8}{16}
  \\ \nonumber
        	&\ + \frac{\sigma^2}{2} \Bigg[   2\conj{x_1}x_2|x_1|^2|x_2|^2    
\\ \nonumber        	
       &\qquad 	+(j+\alpha) \frac{\sigma^2}{2} 	\Big[ 2\conj{x_1}x_2|x_1+x_2|^2
        	+|x_1|^2\left(2\conj{x_1}\conj{x_2}+x_2^2-\conj{x_1}x_2\right) +|x_2|^2\left(2x_1x_2+\conj{x_1}^2-\conj{x_1}x_2\right) \Big] 
        	 \\ \nonumber
  &+(j+\alpha) (j+\alpha+1)\frac{\sigma^4}{4}\Big[
  2|x_1+x_2|^2+\conj{x_1}(2\conj{x_1}+\conj{x_2})+x_2(2x_2+x_1)-|x_1|^2-|x_2|^2  \Big]
    \\ 
        	&\ +2 (j+\alpha) (j+\alpha+1)(j+\alpha+2)\frac{\sigma^6}{8}
  \Bigg]      	
  	\Bigg\}.      	\label{h1}
        	\end{align}
The result  for $k=2$ reads
			 \begin{align} 
\nonumber
&			H_{j-2,j}^{12} = -2\pi \Gamma(j+\alpha)\left( \frac{\sigma^2}{2} \right)^{j+\alpha} \\ \nonumber
			\times &\Bigg\{ \vert x_1 \vert^2 \vert x_2 \vert^2 \left[ (x_2+\conj{x_2} )|x_1|^2+(x_1+\conj{x_1} )|x_2|^2\right]\\
			\nonumber
			&+ (j+\alpha)\frac{\sigma^2}{2}\left[ |x_1+x_2|^2\left((x_2+\conj{x_2} )|x_1|^2+(x_1+\conj{x_1} )|x_2|^2\right)\!+|x_1|^2(x_1x_2^2+\conj{x_1}\,\conj{x_2}^2)+|x_2|^2(x_1^2x_2+\conj{x_1}^2\conj{x_2})\right]\\
			\nonumber
			&+ (j+\alpha)(j+\alpha+1)\frac{\sigma^4}{4}\Big[ |x_1+x_2|^2(x_2+\conj{x_2}+x_1+\conj{x_1})+x_1x_2^2+\conj{x_1}\,\conj{x_2}^2+x_1^2x_2+\conj{x_1}^2\conj{x_2}\Big]\\
			\nonumber
			&+ (j+\alpha)(j+\alpha+1)(j+\alpha+2)\frac{\sigma^6}{8}(x_2+\conj{x_2}+x_1+\conj{x_1})\\
			\nonumber
			&+ \frac{\sigma^2}{2}\Bigg[\conj{x_1}x_2\left((x_2+\conj{x_2} )|x_1|^2+(x_1+\conj{x_1} )|x_2|^2\right)+(\conj{x_1}+x_2)|x_1|^2|x_2|^2\\
			\nonumber
			&\qquad+(j+\alpha)\frac{\sigma^2}{2}\left[ \conj{x_1}x_2(x_2+\conj{x_2}+x_1+\conj{x_1})+(\conj{x_1}+x_2)|x_1+x_2|^2+x_1x_2(x_2+\conj{x_2} )+\conj{x_1}\conj{x_2}(x_1+\conj{x_1} )\right]\\
			&\qquad+(j+\alpha)(j+\alpha+1)\frac{\sigma^4}{4}(2(x_2+\conj{x_1} )+x_1+\conj{x_2} )
			\Bigg]\Bigg\}.       \label{h2}
					\end{align}  
        Similarly, for $k=3$ we compute
         \begin{align}	
        	\nonumber
        H_{j-3,j}^{12} 	=& 2\pi \Gamma(j+\alpha)\left( \frac{\sigma^2}{2} \right)^{j+\alpha} \\ \nonumber
        	 \times \Bigg\{ &\left( |x_1|^2|x_2|^2+ (j+\alpha+1)(j+\alpha)\frac{\sigma^4}{4} \right) 
  \left(\conj{x_1} \conj{x_2}+ {x}_1 x_2+|x_1+x_2|^2\right) 
        	\\ \nonumber
        	&+(j+\alpha)\frac{\sigma^2}{2}\left(|x_1+x_2|^2(\conj{x_1} \conj{x_2}+ {x}_1 x_2)+3|x_1|^2|x_2|^2+(\conj{x_1} x_2+\conj{x_2} x_1)(|x_1|^2+|x_2|^2)\right) 
        	\\ \nonumber
        	&+ \frac{\sigma^2}{2}\Bigg[ 2|x_1|^2|x_2|^2+ (|x_1|^2x_2+|x_2|^2\conj{x_1})(\conj{x_1}+x_2)
\\ 
        	&\qquad + 
(j+\alpha)\frac{\sigma^2}{2}\left(x_1x_2+\conj{x_1} \conj{x_2}+ 2\conj{x_1} x_2+|x_1|^2+|x_2|^2\right)        
        	\Bigg] 
        		\Bigg\}.
        		\label{h3}
        \end{align}        
        In the case $k=4$ we obtain
        	        \begin{align}\label{h4}
        H_{j-4,j}^{12} 	= -2\pi \Gamma(j+\alpha) \left(\frac{\sigma^2}{2}\right)^{j+\alpha} \Bigg\{& 
        	|x_1|^2|x_2|^2(x_1 + \conj{x_1}+x_2+\conj{x_2})
        	  	\\
        	+&  (j+\alpha)\frac{\sigma^2}{2}
        	\left( (x_1 + \conj{x_1})|x_2|^2+ (x_2+\conj{x_2})|x_1|^2\right)
        	+\frac{\sigma^2}{2} \conj{x_1}x_2(x_1+\conj{x_2})
        	\Bigg\}, 
        	\nonumber
        \end{align}
and finally at $k=5$ we have
        \begin{align}
        	H_{j-5,j}^{12} =& 4\pi \int_{0}^\infty \vert x_1 \vert ^2 \vert x_2 \vert^2  r^{2j+2\alpha-1} e^{-2r^2/\sigma^2}\diff r 
        	 =2\pi \Gamma(j+\alpha) \left( \frac{\sigma^2}{2} \right)^{j+\alpha} \vert x_1 \vert ^2 \vert x_2 \vert^2.
        	 \label{h5}
        \end{align}
        
The matrix elements of $H_{ij}^{1\conj 2}$ determining the overlap $\widetilde{\mathcal{O}_N}(x_1,x_2)$ can be obtained by  interchanging $x_2 \leftrightarrow \conj{x_2}$ in the above formulas.

\end{appendix}


\end{document}